\documentclass[12pt, draftclsnofoot, onecolumn]{IEEEtran} 
\usepackage{amsmath}
\usepackage{amssymb}
\usepackage{amsfonts}
\usepackage{graphicx}
\usepackage{epsfig}
\usepackage{subfigure}
\usepackage{psfrag}
\usepackage{cite}
\usepackage{latexsym}
\usepackage{url}
\usepackage{color}
\PassOptionsToPackage{bookmarks={false}}{hyperref}
\IEEEoverridecommandlockouts

\begin{document}
\title{Max-Min Fairness in IRS-Aided Multi-Cell MISO Systems with Joint Transmit and Reflective Beamforming}

\author{Hailiang Xie, Jie Xu, and Ya-Feng Liu\\
\thanks{Part of this paper will be presented in the IEEE International Conference on Communications (ICC), Dublin, Ireland, 2020 \cite{my_conf}.}
\thanks{H. Xie is with the School of Information Engineering, Guangdong University of Technology, Guangzhou 510006, China (e-mail: hailiang.gdut@gmail.com).}
\thanks{J. Xu is with the Future Network of Intelligence Institute (FNii) and the School of Science and Engineering, The Chinese University of Hong Kong (Shenzhen), Shenzhen 518172, China (e-mail: xujie@cuhk.edu.cn). J.~Xu is the corresponding author.}
\thanks{Y.-F. Liu is with the State Key Laboratory of Scientific and Engineering Computing, Institute of Computational Mathematics and Scientific/Engineering Computing, Academy of Mathematics and Systems Science, Chinese Academy of Sciences, Beijing 100190, China (e-mail: yafliu@lsec.cc.ac.cn). }}
\setlength\abovedisplayskip{2.5pt}
\setlength\belowdisplayskip{2.5pt}

\maketitle
\begin{abstract}
This paper investigates an intelligent reflecting surface (IRS)-aided multi-cell multiple-input single-output (MISO) network with a set of multi-antenna base stations (BSs) each communicating with multiple single-antenna users, in which an IRS is dedicatedly deployed for assisting the wireless transmission and suppressing the inter-cell interference. Under this setup, we jointly optimize the coordinated transmit beamforming vectors at the BSs and the reflective beamforming vector (with both reflecting phases and amplitudes) at the IRS, for the purpose of maximizing the minimum weighted signal-to-interference-plus-noise ratio (SINR) at the users, subject to the individual maximum transmit power constraints at the BSs and the reflection constraints at the IRS. To solve the non-convex min-weighted-SINR maximization problem, we first present an {\emph{exact}}-alternating-optimization approach to optimize the transmit and reflective beamforming vectors in an alternating manner, in which the transmit and reflective beamforming optimization subproblems are solved {\emph{exactly}} in each iteration by using the techniques of second-order-cone program (SOCP) and semi-definite relaxation (SDR), respectively. However, the exact-alternating-optimization approach has high computational complexity, and may lead to compromised performance due to the uncertainty of randomization in SDR. To avoid these drawbacks, we further propose an {\emph{inexact}}-alternating-optimization approach, in which the transmit and reflective beamforming optimization subproblems are solved {\emph{inexactly}} in each iteration based on the principle of successive convex approximation (SCA). In addition, to further reduce the computational complexity, we propose a low-complexity inexact-alternating-optimization design, in which the reflective beamforming optimization subproblem is solved more {\emph{inexactly}}. Via numerical results, it is shown that the proposed three designs achieve significantly increased min-weighted-SINR values, as compared with benchmark schemes without the IRS or with random reflective beamforming. It is also shown that the inexact-alternating-optimization design outperforms the exact-alternating-optimization one in terms of both the achieved min-weighted-SINR value and the computational complexity, while the low-complexity inexact-alternating-optimization design has much lower computational complexity with slightly compromised performance. Furthermore, we show that our proposed design can be applied to the scenario with unit-amplitude reflection constraints, with a negligible performance loss.
\end{abstract}


\newtheorem{definition}{\underline{Definition}}[section]
\newtheorem{fact}{Fact}
\newtheorem{assumption}{Assumption}
\newtheorem{theorem}{\underline{Theorem}}[section]
\newtheorem{lemma}{\underline{Lemma}}[section]
\newtheorem{corollary}{\underline{Corollary}}[section]
\newtheorem{proposition}{\underline{Proposition}}[section]
\newtheorem{example}{\underline{Example}}[section]
\newtheorem{remark}{\underline{Remark}}[section]
\newtheorem{algorithm}{\underline{Algorithm}}[section]
\newtheorem{proof}{Proof}[section]
\newcommand{\mv}[1]{\mbox{\boldmath{$ #1 $}}}

\section{Introduction}

Recent technical advancements in Internet of things (IoT) and artificial intelligence (AI) are expected to enable various new applications such as autonomous driving, augmented reality (AR), virtual reality (VR), and industrial automation. To make the IoT and AI vision a reality, the cellular networks are evolving towards the fifth generation (5G) and beyond to support massive wireless devices with diverse quality of service (QoS) requirements, such as significantly increased spectrum efficiency, ultra-low transmission latency, and  extremely-high communication reliability \cite{5G,6G}. Towards this end, small base stations (BSs) are densely deployed to shorten the distances with end users \cite{small_cells1,small_cells2,small_cells3}, and device-to-device (D2D) communications are enabled underlying conventional cellular transmissions to create more spectrum reuse opportunities \cite{D2D_survey,D2D_2011,D2D_2014}. However, the emergence of small BSs and D2D communications in 5G-and-beyond cellular networks also introduces severe co-channel interference among different cells and different D2D links, which needs to be dealt with carefully. In the literature, various approaches have been proposed to mitigate or even utilize the co-channel interference, some examples including coordinated transmit/receive beamforming \cite{Co_beam_2010, Co_beam_2011, ICIC, Co_beam_2013} and network multiple-input multiple-output (MIMO) \cite{Net_MIMO_2006, Net_MIMO_2009, Net_MIMO_2010, Net_MIMO_2016}. For instance, in the coordinated beamforming, different BSs are enabled to share their channel state information (CSI) in order to design their transmit/receive beamfoming vectors in a coordinated manner to mitigate the inter-cell co-channel interference.

Recently, intelligent reflecting surface (IRS) has emerged as another promising technology for beyond-5G cellular networks\cite{IRS_wu,IRS_survey,LISA,Emil}. IRS is a passive meta-material panel consisting of a large number of reflecting units, each of which can introduce an independent phase shift on radio-frequency (RF) signals to facilitate the wireless transmission. In particular, by jointly controlling these phase shifts, the IRS can form reflective signal beamforming (to combat the severe path loss), such that the reflected signals can be coherently combined with the directly transmitted signals at intended receivers for enhancing the desirable signal strength, or destructively combined at unintended receivers for suppressing the undesirable interference. As the IRS is a passive device with no dedicated power consumption, it is envisioned as a green and cost-effective solution to enhance both the spectrum- and energy-efficiency of future cellular networks\cite{IRS_wu,IRS_survey,LISA,Emil}. It is also envisioned that the IRS can be a viable new solution to help enhance the performance of interfering wireless networks by reconfigurating the wireless transmission environment.

How to jointly design the transmit beamforming at wireless transmitters (e.g., BSs) and the reflective beamforming at the IRS is one of the key issues to be tackled in IRS-aided wireless communication systems. In the literature, there have been several prior works \cite{IRS_single1,IRS_single2,IRS_multiuser,IRS_NOMA1,IRS_NOMA2,IRS_OFDM,IRS_OFDM_MU,IRS_Capacity,SWIPT1,SWIPT2,SWIPT3,IRS_multicell} investigating this problem under different setups. For instance, the authors in \cite{IRS_single1,IRS_single2} aimed to maximize the received signal-to-noise ratio (SNR) in a point-to-point IRS-aided multiple-input single-output (MISO) communication system, for which the techniques of semi-definite relaxation (SDR) \cite{IRS_single1} and manifold optimization \cite{IRS_single2} are employed, respectively. \cite{IRS_Capacity} characterized the fundamental capacity limit of the IRS-aided point-to-point MIMO system. Furthermore, \cite{IRS_multiuser} considered the signal-to-interference-plus-noise ratio (SINR)-constrained power minimization in IRS-aided multiuser MISO downlink communication systems, in which the alternating optimization is employed to update the transmit and reflective beamforming vectors in an alternating manner, and the SDR is employed to optimize the reflective beamforming vector. In addition,  \cite{IRS_OFDM,IRS_OFDM_MU} studied the IRS-aided orthogonal frequency division multiplexing (OFDM) systems, in which the reflective beamforming vector must be designed over all subcarriers while the transmit beamforming vectors can be designed independently over each subcarrier. \cite{IRS_OFDM} aimed to maximize the achievable rate for an IRS-aided single-input single-output (SISO) OFDM system, and \cite{IRS_OFDM_MU} investigated the average sum-rate maximization problem in an IRS-aided multiuser MISO OFDM system. Moreover, the IRS has also been employed under other communication setups, such as non-orthogonal multiple access (NOMA) \cite{IRS_NOMA1,IRS_NOMA2}, and simultaneous wireless information and power transfer (SWIPT) systems\cite{SWIPT1,SWIPT2,SWIPT3}. Nevertheless, all the above prior works \cite{IRS_single1,IRS_single2,IRS_multiuser,IRS_NOMA1,IRS_NOMA2,IRS_OFDM,IRS_OFDM_MU,IRS_Capacity,SWIPT1,SWIPT2,SWIPT3} focused on a single-cell setup. This thus motivates us to use IRS to facilitate the interfering multi-cell communications in this work.

\begin{figure}
\centering
 \epsfxsize=1\linewidth
    \includegraphics[width=9cm]{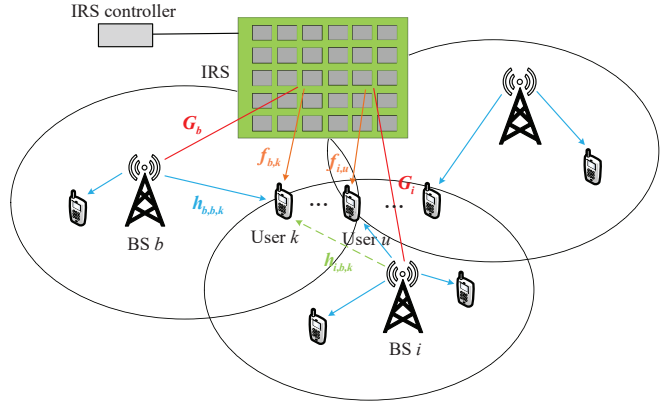}
\caption{Illustration of an IRS-aided multi-cell MISO system.} \label{fig:Multicell}
\vspace{-1em}
\end{figure}

In this paper, we consider an IRS-aided multi-cell MISO system as shown in Fig.~\ref{fig:Multicell}, where an IRS is dedicatedly deployed at the cell boundary to assist the wireless transmission from (small) BSs to users and suppress their inter-cell interference. We assume that there is one multi-antenna BS serving multiple single-antenna users in each cell. Under this setup, the main results of this paper are listed as follows. \begin{itemize}
	\item Our objective is to jointly optimize the coordinated transmit beamforming vectors at the multiple BSs and the reflective beamforming vector at the IRS, to maximize the minimum weighted received SINR at users, subject to the individual maximum transmit power constraints at the BSs, and the reflection constraints at the IRS. However, due to the coupling between the transmit and reflective beamforming vectors, the formulated minimum SINR maximization problem is highly non-convex and thus difficult to be optimally solved.
	\item To solve the non-convex minimum SINR maximization problem, we first present an exact-alternating-optimization approach to optimize the transmit and reflective beamforming vectors in an alternating manner. In each iteration, we solve the transmit and reflective beamforming optimization subproblems {\it exactly} by handling a series of feasibility second-order cone programs (SOCPs) together with a bisection search, and by using the SDR technique, respectively. However, the exact-alternating-optimization approach is with high computational complexity and may lead to compromised performance, due to the uncertainty of randomizations in SDR.
	\item To avoid the above drawbacks, we further propose an inexact-alternating-optimization approach, where in each iteration, the transmit and reflective beamforming optimization subproblems are solved {\it inexactly} based on the principle of successive convex approximation (SCA). Specifically, in the inexact-alternating-optimization approach, we only need to find approximate solutions to the transmit/reflective beamforming subproblems with an increased min-weighted-SINR value at each iteration (instead of exactly solving them with optimal/converged solutions), thus leading to reduced computational complexity and guaranteed performance.
	\item In addition, to further reduce the computational complexity, we propose another low-complexity inexact-alternating-optimization design. In each iteration, we update the reflective beamforming more inexactly by using the subgradient projection method.
	\item Finally, we present numerical results to validate the performance of our proposed approaches. It is shown that the proposed three designs achieve significantly increased min-weighted-SINR values, as compared with benchmark schemes without the IRS or with random reflective beamforming. It is also shown that the inexact-alternating-optimization design outperforms the exact-alternating-optimization one in terms of both the achieved min-weighted-SINR value and the computational complexity, while the low-complexity inexact-alternating-optimization design has much lower computational complexity with slightly compromised performance. Furthermore, we show that our proposed design can be applied to the scenario with unit-amplitude reflection constraints, with a negligible performance loss.
\end{itemize}

It is worth noting that there is only one existing work \cite{IRS_multicell} that studied the weighted sum-rate maximization in IRS-aided multi-cell networks by applying the block coordinate descent algorithm together with the majorization minimization and the complex circle manifold. Nevertheless, this paper is different from \cite{IRS_multicell} in the following two aspects. First, while \cite{IRS_multicell} focsed on the weighted sum-rate maximization, this paper considers a different objective of the min-weighted-SINR maximization with distinct solution approaches. Second, while \cite{IRS_multicell} only optimized the reflection phases at the IRS by considering unit amplitudes, this paper further exploits the optimization of reflection amplitudes to enhance the communication performance.

The remainder of this paper is organized as follows. Section II introduces the IRS-aided multi-cell MISO system model and presents the min-weighted-SINR maximization problem of interest. Sections III-V propose three different approaches to solve the formulated problem, namely exact-alternating-optimization, inexact-alternating-optimization, and low-complexity inexact-alternating-optimization, respectively. Section VI presents numerical results to evaluate the performance of the proposed approaches. Finally, Section VII concludes this paper with future research directions.

{\it Notations:} Boldface letters refer to vectors (lower  case) or matrices (upper case). For a square matrix $\mv{S}$, ${\mathtt{Tr}}(\mv{S})$ denotes its trace, while $\mv{S}\succeq \mv{0}$ and $\mv{S}\preceq \mv{0}$ mean that $\mv{S}$ is positive and negative semidefinite, respectively. For an arbitrary-size matrix $\mv{M}$, ${\mathrm{rank}}(\mv{M})$, $\mathrm{vec}(\mv{M})$, $\mv{M}^H$, and $\mv{M}^T$ denote its rank, vectorization, conjugate transpose, and transpose, respectively, and $[\mv{M}]_{ik}$ denotes the element in the $i$-th row and $k$-th column of $\mv{M}$. $\mv{I}$, $\mv{0}$, and $\mv e_i$ denote an identity matrix, an all-zero matrix and a vector with the $i$-th element being one and others being zero, respectively, with appropriate dimensions. The distribution of a circularly symmetric complex Gaussian (CSCG) random vector with mean vector $\mv{x}$ and covariance matrix $\mv{\Sigma}$ is denoted by $\mathcal{CN}(\mv{x,\Sigma})$; and $\sim$ stands for ``distributed as''. $\mathbb{C}^{x\times y}$ denotes the space of $x\times y$ complex matrices. $\mathbb{R}$ denotes the set of real numbers. ${\mathbb{E}}(\cdot)$ denotes the stochastic expectation. $\|\mv{x}\|$ denotes the Euclidean norm of a complex vector $\mv{x}$, and $|z|$ denotes the magnitude of a complex number $z$. $[\mv{x}]_n$ denotes the $n$-th element of $\mv{x}$. $\mathrm{arg}(x)$ denotes the phase of a complex number $x$. $\mathrm{diag}(a_1,\ldots, a_N)$ denotes a diagonal matrix with the diagonal elements being $a_1,\ldots, a_N$. $\mathrm{Re}(x)$ and $\mathrm{Im}(x)$ denote the real and imaginary parts of a complex number $x$, respectively. $\mathrm{\mv{Conv}}$ $(\mathcal C)$ denotes the convex hull of a set $\mathcal C$. $\nabla f(\mv{x})$ denotes the gradient vector of function $f(\mv{x})$.

\section{System Model and Problem Formulation}

As shown in Fig. \ref{fig:Multicell}, we consider an IRS-aided multi-cell MISO system, where an IRS is dedicatedly deployed at the cell boundary to assist the multi-cell communication and suppress the inter-cell interference, especially for cell-edge users. Suppose that there are $B$ (small) cells (or BSs) in this system and the IRS consists of $N>1$ reflecting units. Let $\mathcal B \triangleq \{1,...,B\}$ and $\mathcal N  \triangleq \{1,..., N\}$ denote the set of BSs in the system and that of reflecting units at the IRS, respectively. In each cell $b\in\mathcal B$, the BS with $M\ge1$ antennas communicates with $K_b\ge 1$ single-antenna users, where $\mathcal K_b \triangleq\{1,\ldots,K_b\}$ denotes the set of users in cell $b$. Accordingly, the total number of users in the system is denoted as $K=\sum_{b\in\mathcal{B}}K_b$. The IRS can adaptively adjust the reflecting phases and amplitudes\footnote{Note that there have been various methods to adaptively control the amplitude of each reflecting unit, together with its phase (see, e.g., \cite{amplitude1,amplitude2,amplitude3}). For instance, it can be practically realized by dynamically changing the connected resistor load \cite{amplitude3}. In the literature, such joint phase and amplitude control at the IRS has been widely adopted in prior works (see, e.g., \cite{IRS_est_OFDM,amp_phase}), which generally serves as a performance upper bound for designs with phase control only (with the reflecting amplitude fixed to be 1). Also note that our proposed joint phase and amplitude control at the IRS can be extended to the case with phase control only, with a slightly compromised performance, as will be shown in Section VI (see Fig. \ref{fig:antenna_number_unit_amplitude} further ahead).} to form reflective signal beam, such that the reflected signal can be coherently combined with the directly transmitted signal at the intended user or destructively combined at the unintended users.

We consider quasi-static narrow-band channel models, where the wireless channels remain unchanged within each transmission block of our interest but may change over different blocks. To help characterize the fundamental performance upper bound for gaining insights, we assume that the perfect CSI of all involved channels is known at both the BSs and the IRS controller to facilitate the joint transmit and reflective beamforming design\footnote{To obtain the global CSI, there is a central node to coordinate channel estimation at each cell and collect all the CSI to coordinate the joint design. In practice, due to the passive nature of the IRS (without active RF chains), the conventional channel estimation methods are not applicable to acquire the CSI associated with the IRS. To tackle this issue, several new channel estimation methods have been proposed in the literature. For instance, \cite{IRS_est_OFDM} proposed to estimate the IRS-related wireless channels by sequentially turning on each reflecting unit (with the other units being off), while \cite{IRS_est_MU} proposed a three-phase channel estimation framework in the IRS-aided uplink multiuser system. Notice that in this paper, the overhead required for channel estimation, CSI feedback to the central node, and the signaling between the central node and the BSs/IRS have not been taken into account, for the purpose of characterizing the performance upper bound to gain the fundamental system design insight. 
}. Let $\mv{G}_b\in {\mathbb C}^{N\times M}$ denote the channel matrix from BS $b$ to the IRS, $\mv{f}_{b,k}\in {\mathbb C}^{N\times 1}$ denote the channel vector from the IRS to user $k$ in cell $b$, and $\mv{h}_{i,b,k}\in {\mathbb C}^{M\times 1}$ denote the channel vector from BS $i$ to user $k$ in cell $b$. Let $s_{b,k}$ denote the desirable message for user $k\in\mathcal K_b$ from BS $b$ and $\mv w_{b,k}\in\mathbb C^{M\times1}$ denote the corresponding transmit beamforming vector by BS $b$, where $s_{b,k}$'s are assumed to be independent and identically distributed (i.i.d.) CSCG random variables with zero mean and unit variance, i.e., $s_{b,k}\sim\mathcal{CN}(0,1)$. Accordingly, the transmitted signal by BS $b\in\mathcal B$ is given by 
\begin{align}
\mv x_b = \sum_{k\in\mathcal{K}_b}\mv w_{b,k} s_{b,k}.\label{transm_signal}
\end{align}
Suppose that each BS has a maximum transmit power budget denoted by $P_b$. Then we have $\mathbb{E}(\|\mv x_b\|^2) = \sum_{k\in\mathcal{K}_b}\|\mv w_{b,k}\|^2 \le P_b, \forall b\in\mathcal B$.

As for the reflection at the IRS, let $\theta_n\in[0, 2\pi)$ and $\beta_n\in[0, 1]$ denote the phase shift and the reflection amplitude imposed by the $n$-th reflecting unit on the incident signal\cite{IRS_wu}, respectively. Accordingly, let $\mv{\Theta}=\mathrm{diag}\left(\beta_1 e^{j\theta_{1}},\ldots,\beta_N e^{j\theta_{N}}\right) $ represent the reflection coefficient matrix at the IRS, where $j\triangleq \sqrt{-1}$. Furthermore, let $\mv v=[\beta_1 e^{j\theta_{1}},\ldots,\beta_N e^{j\theta_{N}}]^H$ denote the reflective beamforming vector, where each element $n$, denoted by $v_n$, must satisfy $|v_n|\!\le\!1$\footnote{Notice that different from the conventional active relay, the considered IRS (even with amplitude control) is generally passive without dedicated transmitter modules equipped, and thus is more energy efficient\cite{comp_relay1}. Besides, while relays are normally operated in the half-duplex (HD) mode to avoid severe self-interference, the IRS can be implemented like the full-duplex (FD) mode without self-interference caused therein.}, $\forall n\in\mathcal N$. As a consequence, we have the combined reflective channel from BS $i$ to user $k$ in cell $b$ as $\mv{f}^H_{b,k}\mv{\Theta}\mv{G}_i = \mv v^H\mv\Phi_{i,b,k}$, where $\mv\Phi_{i,b,k} = \mathrm{diag}(\mv{f}^H_{b,k})\mv{G}_i$. Notice that this transformation separates the reflective beamforming vector $\mv v$ from the reflective channels, which will significantly facilitate our derivation later. By combining the directly transmitted and reflected signals, the signal received at user $k$ in cell $b$ is accordingly expressed as
\begin{align}
y_{b,k}=&(\mv v^H\mv\Phi_{b,b,k}+\mv h^H_{b,b,k})\mv w_{b,k} s_{b,k}+\sum_{u\in \mathcal{K}_b, u\neq k} (\mv v^H\mv\Phi_{b,b,k}+\mv h^H_{b,b,k}) \mv w_{b,u} s_{b,u} \nonumber\\
&+\sum\limits_{i\in\mathcal B, i\neq b}\sum_{u\in\mathcal{K}_i}( \mv v^H\mv\Phi_{i,b,k}+ \mv h^H_{i,b,k})\mv w_{i,u} s_{i,u} + n_{b,k}, \label{fm:1}
\end{align}
where $n_{b,k}$ denotes the additive white Gaussian noise (AWGN) at the receiver of user $k$ in cell $b$ with zero mean and variance $\sigma_{b,k}^2$, i.e., $n_{b,k}\sim\mathcal{CN}(0,\sigma_{b,k}^2), \forall k\in\mathcal K_b, b\in\mathcal B$. By treating the interference as noise, the received SINR at user $k$ in cell $b$ is given by
\begin{align}
&\mathrm{\gamma}_{b,k}(\mv v, \{\mv w_{i,u}\})\!=\!\frac{|(\mv v^H\mv\Phi_{b,b,k}+ \mv h^H_{b,b,k})\mv w_{b,k}|^2}{\sum\limits_{u\in \mathcal{K}_b, u\neq k}|(\mv v^H\mv\Phi_{b,b,k}\!+\!\mv h^H_{b,b,k})\mv w_{b,u}|^2\!+\!\sum\limits_{i\in\mathcal B, i\neq b}\sum\limits_{u\in\mathcal{K}_i}|(\mv v^H\mv\Phi_{i,b,k}+ \mv h^H_{i,b,k})\mv w_{i,u}|^2+\sigma_{b,k}^2},\nonumber\\
&~~~~~~~~~~~~~~~~~~~~\forall k\in\mathcal{K}_b, b\in\mathcal B.\label{fm:2}
\end{align}

Our objective is to maximize the users' communication performance in a fair manner as those in\cite{Co_beam_2013,User_fair1}. As a result, we consider the max-min fairness problem with the objective of maximizing the minimum weighted SINR of all users, by jointly optimizing the transmit beamforming vectors $\{\mv w_{b,k}\}$ at the BSs and the reflective beamforming vector $\mv v$ at the IRS, subject to the individual transmit power constraints at the BSs and the reflection constraints at the IRS. Let $\alpha_{b,k}\!>\!0$ denote a weight parameter for user $k\in\mathcal K_b$ in cell $b$ to characterize the fairness among all users, where a larger value of $\alpha_{b,k}$ indicates that user $k$ in cell $b$ has a higher priority in transmission. Therefore, the min-weighted-SINR maximization problem is formulated as
\begin{align}
\mathtt{(P1)}:&\mathop\mathtt{max}_{\mv v,\{\mv w_{b,k}\}}~\mathop\mathtt{min}_{k\in\mathcal{K}_b, b\in\mathcal B}~\frac{\mathrm{\gamma}_{b,k}(\mv v, \{\mv w_{b,k}\})}{\alpha_{b,k}} \label{Problem:ori:1} \\
&~~~{\mathtt{s.t.}}~~~~~~\sum_{k\in\mathcal{K}_b}\|\mv w_{b,k}\|^2 \le P_b, \forall b\in\mathcal B \label{Problem:ori:2} \\
&~~~~~~~~~~~~~~|v_n|\le 1, \forall n\in\mathcal N. \label{Problem:ori:3}
\end{align}
To facilitate the derivation, we first introduce an auxiliary variable $t$ and reformulate problem (P1) as the following equivalent problem:
\begin{align}
&\mathtt{(P1.1)}:\mathop\mathtt{max}_{\mv v,\{\mv w_{b,k}\},t}~ t \nonumber \\
&~~~~~~~~~~~~~{\mathtt{s.t.}}~~~\mathrm{\gamma}_{b,k}(\mv v, \{\mv w_{b,k}\})\ge\alpha_{b,k} t, \forall k\in\mathcal{K}_b, b\in\mathcal B \label{Problem:ori_epi:1} \\
&~~~~~~~~~~~~~~~~~~~~\sum_{k\in\mathcal{K}_b}\|\mv w_{b,k}\|^2 \le P_b, \forall b\in\mathcal B \label{Problem:ori_epi:2} \\
&~~~~~~~~~~~~~~~~~~~~~|v_n|\le 1, \forall n\in\mathcal N. \label{Problem:ori_epi:3}
\end{align}
Notice that the formulated problem (P1)/(P1.1) is quite different and more challenging to tackle compared to the conventional max-min fairness coordinated beamforming problems \cite{Co_beam_2010,Co_beam_2011}. Specifically, conventional max-min fairness coordinated beamforming problems in the MISO system are essentially convex, which have been well studied with optimal solution \cite{Co_beam_2011}. However, in this paper, the existence of the IRS not only introduces additional IRS constraints, but also makes the reflective beamforming vector highly coupled with the coordinated transmit beamforming vectors at the SINR terms, thus making the formulated problem non-convex and highly nonlinear. So far, there is no efficient way to address it yet. Therefore, the existence of the IRS makes the problem much more challenging to solve.

\section{Exact Alternating Optimization}\label{sec:III}
In this section, we propose an exact-alternating-optimization approach to solve the difficult non-convex min-weighted-SINR maximization problem (P1), in which the transmit beamforming vectors $\{\mv w_{b,k}\}$ and the reflective beamforming vector $\mv v$ are optimized in an alternating manner, with the other being fixed. For notational convenience, suppose that at each iteration $l \ge 0$, the obtained beamforming vectors are denoted by $\{\mv w^{(l)}_{b,k}\}$ and $\mv v^{(l)}$, where $\mv v^{(0)}$ and $\{\mv w^{(0)}_{b,k}\}$ denote the initial beamforming vectors. Notice that by ``exact'', we mean that at each iteration, the beamforming vectors $\{\mv w_{b,k}^{(l)}\}$ and $\mv v^{(l)}$ are obtained by exactly solving the corresponding transmit and reflective beamforming optimization subproblems, respectively. 

\subsection{Coordinated Transmit Beamforming Optimization}
First, we present the coordinated transmit beamforming design under any given reflective beamforming vector $\mv v$. For notational convenience, we define $\mv a_{i,b,k}=\mv\Phi_{i,b,k}^H\mv v + \mv h_{i,b,k}$ as the effective or combined channel from BS $i$ to user $k$ in cell $b, \forall k\in\mathcal{K}_b, i,b\in\mathcal{B}$. Accordingly, the coordinated transmit beamforming optimization problem becomes
\begin{align}
\mathrm{(P2):}&\mathop\mathtt{max}_{\{\mv w_{b,k}\},t} ~ t \nonumber \\
&~~~{\mathtt{s.t.}}
~~\frac{|\mv a^H_{b,b,k}\mv w_{b,k}|^2}{\sum\limits_{u\in\mathcal{K}_b, u\neq k}|\mv a^H_{b,b,k}\mv w_{b,u}|^2\!+\!\sum\limits_{i\in\mathcal B, i\neq b}\sum\limits_{u\in\mathcal{K}_i}|\mv a^H_{i,b,k}\mv w_{i,u}|^2\!+\!\sigma_{b,k}^2}\!\ge\!\alpha_{b,k} t, \forall k\in\mathcal{K}_b, b\in\mathcal B \label{Problem:given_phase:1} \\
&~~~~~~~~~\sum_{k\in\mathcal{K}_b}\|\mv w_{b,k}\|^2 \le P_b, \forall b\in\mathcal B. \label{Problem:given_phase:2}
\end{align}
It is observed that problem (P2) is still non-convex. To tackle this issue, we introduce the following feasibility problem (P2.1), which is obtained based on problem (P2) by fixing $t$.
\begin{align}
\mathrm{(P2.1):}&\mathop\mathtt{find}~ \{\mv w_{b,k}\} \nonumber \\
&~~{\mathtt{s.t.}}
~\frac{|\mv a^H_{b,b,k}\mv w_{b,k}|^2}{\sum\limits_{u\in\mathcal{K}_b, u\neq k}|\mv a^H_{b,b,k}\mv w_{b,u}|^2\!+\!\sum\limits_{i\in\mathcal B, i\neq b}\sum\limits_{u\in\mathcal{K}_i}|\mv a^H_{i,b,k}\mv w_{i,u}|^2\!+\!\sigma_{b,k}^2}\!\ge\!\alpha_{b,k} t, \forall k\in\mathcal{K}_b, b\in\mathcal B \label{Problem:given_phase_f:1} \\
&~~~~~~~~\sum_{k\in\mathcal{K}_b}\|\mv w_{b,k}\|^2 \le P_b, \forall b\in\mathcal B. \label{Problem:given_phase_f:2}
\end{align}
In particular, suppose that the optimal solution of $t$ to problem (P2) is given by $t^\star$. It is thus clear that if problem (P2.1) is feasible under any given $t$, then we have $t \le t^\star$; while if (P2.1) is infeasible, then it follows that $t > t^\star$. Therefore, problem (P2) can be equivalently solved by checking the feasibility of problem (P2.1) under any given $t > 0$, together with a bisection search over $t > 0$.

Therefore, to solve problem (P2), we only need to solve problem (P2.1) under any fixed $t > 0$, by using SOCP as follows \cite{SOCP}. Towards this end, we notice that the SINR constraints in (\ref{Problem:given_phase:1}) can be reformulated as
\begin{align}
\big(1+\frac{1}{\alpha_{b,k} t}\big)|\mv a^H_{b,b,k}\mv w_{b,k}|^2 \ge \sum\limits_{i\in\mathcal B}\sum\limits_{u\in\mathcal{K}_i}|\mv a^H_{i,b,k}\mv w_{i,u}|^2+\sigma_{b,k}^2, \forall k\in\mathcal{K}_b, b\in\mathcal B. \label{tf_socp:1}
\end{align}
Based on (\ref{tf_socp:1}), it is evident that if $\{\mv w_{b,k}\}$ is a feasible solution to problem (P2.1), then any phase rotation of $\{\mv w_{b,k}\}$ will still be feasible. Without loss of optimality, we choose the solution of $\{\mv w_{b,k}\}$ such that $\mv a^H_{b,b,k}\mv w_{b,k}$ becomes a non-negative value for any user $k\in\mathcal{K}_b$ in cell $b\in\mathcal{B}$. As a result, we have the following constraints:
\begin{align}
\mv a^H_{b,b,k}\mv w_{b,k} \ge 0, \forall k\in\mathcal{K}_b, b\in\mathcal B,\label{tf_socp:2}
\end{align}
where $\mv a^H_{b,b,k}\mv w_{b,k}$ has a non-negative real part and a zero imaginary part, i.e., $\mathrm{Re}(\mv a_{b,b,k}^H \mv w_{b,k}) \ge 0$ and $\mathrm{Im}(\mv a^H_{b,b,k}\mv w_{b,k})=0$. Accordingly, (\ref{tf_socp:1}) can be further re-expressed as
\begin{align}
\sqrt{1+\frac{1}{\alpha_{b,k} t}}\mv a^H_{b,b,k}\mv w_{b,k} \ge \begin{Vmatrix}
\mv A^H \mv e_{m_{b,k}} \\
\sigma_{b,k}
\end{Vmatrix}_2, \forall k\in\mathcal{K}_b, b\in\mathcal B, \label{tf_socp:3}
\end{align}
where $m_{b,k}=K_1+\dots+K_{b-1}+k,\forall k\in\mathcal{K}_b, b\in\mathcal B$ and $\mv A\in\mathbb C^{K\times K}$ denotes a matrix with the element in its $m_{b,k}$-th row and $m_{i,u}$-th column being $\mv a^H_{i,b,k}\mv w_{i,u}$. Therefore, problem (P2.1) is reformulated as the following equivalent form:
\begin{align}
\mathrm{(P2.2):}\mathop\mathtt{find}~& \{\mv w_{b,k}\} \nonumber \\
{\mathtt{s.t.}}
~&  \|\mathrm{vec}(\mv{W}_b)\|_2 \le \sqrt{P_b}, \forall b\in\mathcal B\\
~&(\ref{tf_socp:2})~\mathrm{and}~(\ref{tf_socp:3}), \nonumber
\end{align}
where $\mv{W}_b=[\mv{w}_{b,1},\dots,\mv{w}_{b,K_b}]$ denotes the matrix consisting of the transmit beamforming vectors at BS $b\in\mathcal{B}$. Problem (P2.2) is an SOCP that can be optimally solved by standard convex optimization solvers such as CVX \cite{CVX}. Therefore, the optimal coordinated transmit beamforming solution to problem (P2) is finally obtained.

\subsection{Reflective Beamforming Optimization}
Next, we optimize the reflective beamforming vector $\mv v$ under any given transmit beamforming $\{\mv w_{b,k}\}$. For notational convenience, we define $\mv c_{i,u,b,k} = \mv\Phi_{i,b,k}\mv w_{i,u}$ and $d_{i,u,b,k} = \mv h^H_{i,b,k}\mv w_{i,u}, \forall k\in\mathcal{K}_b, u\in\mathcal{K}_i, b, i\in\mathcal B$. Then, we have
\begin{align}
|(\mv v^H\mv{\Phi}_{i,b,k} + \mv h^H_{i,b,k})\mv w_{i,u}|^2 = \mv v^H\mv C_{i,u,b,k}\mv v + 2\mathrm{Re}\{\mv v^H\mv u_{i,u,b,k}\} + |d_{i,u,b,k}|^2, \label{fm:3}
\end{align}
where $\mv C_{i,u,b,k}=\mv c_{i,u,b,k}\mv c^H_{i,u,b,k}$ and $\mv u_{i,u,b,k}=\mv c_{i,u,b,k}d_{i,u,b,k}^H$.
Accordingly, the reflective beamforming optimization problem is given by
\begin{align}
&\mathtt{(P3):}~\mathop\mathtt{max}_{\mv v,t}~~~t \nonumber \\
&~~~~~~~~~{\mathtt{s.t.}}~~\frac{\mv v^H\mv C_{b,k,b,k}\mv v + 2\mathrm{Re}\{\mv v^H\mv u_{b,k,b,k}\} + |d_{b,k,b,k}|^2}{\sum\limits_{i\in\mathcal B}\sum\limits_{u\ne k, u\in\mathcal{K}_i}\mv v^H\mv C_{i,u,b,k}\mv v+ 2\mathrm{Re}\{\mv v^H\mv u_{i,u,b,k}\} + |d_{i,u,b,k}|^2 + \sigma_{b,k}^2}\nonumber\\
&~~~~~~~~~~~~~~~\ge\alpha_{b,k} t, \forall k\in\mathcal{K}_b, b\in\mathcal B \label{Problem:given_beam:1} \\
&~~~~~~~~~~~~~~~~|v_n|\le 1, \forall n\in\mathcal N. \label{Problem:given_beam:7}
\end{align}
Notice that problem (P3) is also a non-convex optimization problem. Motivated by the wide application of SDR in solving reflective beamforming optimization problems (see, e.g., \cite{IRS_multiuser}),  we use the well-established SDR technique to solve problem (P3). Towards this end, we first define $|(\mv{v}^H\mv{\Phi}_{i,b,k} + \mv h^H_{i,b,k})\mv w_k|^2 = \bar{\mv v}^{\rm{H}}{\mv R}_{i,u,b,k}\bar{\mv v} + |d_{i,u,b,k}|^2$, where
\begin{align}
{\mv R}_{i,u,b,k} = \begin{bmatrix}
\mv C_{i,u,b,k} & \mv u_{i,u,b,k} \\
\mv u^H_{i,u,b,k} & 0
\end{bmatrix} ~\mathrm{and} ~\bar{\mv v} = \begin{bmatrix}
\mv{v} \\ 1 \end{bmatrix}. \label{R_v}
\end{align}
Accordingly, problem (P3) is re-expressed as
\begin{align}
&\mathtt{(P3.1):}~\mathop\mathtt{max}_{\bar{\mv v},t}~~t \nonumber \\
&~~~~~~~~~~~{\mathtt{s.t.}}
~\frac{\bar{\mv v}^{H}{\mv R}_{b,k,b,k}\bar{\mv v} + |d_{b,k,b,k}|^2}{\sum\limits_{u\in\mathcal{K}_b, u\neq k}\bar{\mv v}^{H}{\mv R}_{b,u,b,k}\bar{\mv v} + |d_{b,u,b,k}|^2 +\sum\limits_{i\in\mathcal B, i\neq b}\sum\limits_{u\in\mathcal{K}_i}\bar{\mv v}^{H}{\mv R}_{i,u,b,k}\bar{\mv v} + |d_{i,u,b,k}|^2 +\sigma_i^2}\nonumber\\
&~~~~~~~~~~~~~~~~~\ge\alpha_{b,k} t, \forall k\in\mathcal{K}_b, b\in\mathcal B \label{Problem:given_beam:2} \\
&~~~~~~~~~~~~~~~~~|\bar v_n|\le 1, \forall n\in\mathcal N \label{Problem:given_beam:3}\\
&~~~~~~~~~~~~~~~~~|\bar v_{N+1}| = 1. \label{Problem:given_beam:4}
\end{align}
Furthermore, we define $\mv V= \bar{\mv v}\bar{\mv v}^H$ with $\mv V \succeq 0$ and $\mathrm{rank}(\mv V)\le 1$. Then problem (P3.1) or (P3) is further reformulated as the following equivalent form:
\begin{align}
&\mathtt{(P3.2):}~\mathop\mathtt{max}_{\mv V,t} ~~ t \nonumber \\
&~~~~~~~~~~~{\mathtt{s.t.}}~~\frac{\mathrm{Tr}({\mv R_{b,k,b,k}}\mv V) + |d_{b,k,b,k}|^2}{\sum\limits_{u\in\mathcal{K}_b, u\neq k}\mathrm{Tr}({\mv R_{b,u,b,k}}\mv V) + |d_{b,u,b,k}|^2 +\sum\limits_{i\in\mathcal B, i\neq b}\sum\limits_{u\in\mathcal{K}_i}\mathrm{Tr}({\mv R_{i,u,b,k}}\mv V) + |d_{i,u,b,k}|^2 + \sigma^2_{b,k}}\nonumber\\
&~~~~~~~~~~~~~~~~~\ge\alpha_{b,k} t, \forall k\in\mathcal{K}_b, b\in\mathcal B \label{Problem:given_beam_SDR:1}\\
&~~~~~~~~~~~~~~~~~ V_{n,n}\le 1, \forall n\in\mathcal N \label{Problem:given_beam_SDR:2}\\
&~~~~~~~~~~~~~~~~~ V_{N+1,N+1}=1 \label{Problem:given_beam_SDR:3}\\
&~~~~~~~~~~~~~~~~~\mv V\succeq 0 \label{Problem:given_beam_SDR:4}\\
&~~~~~~~~~~~~~~~~~\mathrm{rank}(\mv V)\le 1. \label{Problem:given_beam_SDR:5}
\end{align}
However, problem (P3.2) is still challenging to be optimally solved due to the non-convex rank-one constraint in (\ref{Problem:given_beam_SDR:5}). To tackle this issue, we relax this constraint, and obtain a relaxed version of (P3.2) as
\begin{align}
&\mathtt{(P3.3):}\mathop\mathtt{max}_{\mv V,t} ~ t \nonumber \\
&~~~~~~~~~~{\mathtt{s.t.}}~(\ref{Problem:given_beam_SDR:1}),(\ref{Problem:given_beam_SDR:2}), (\ref{Problem:given_beam_SDR:3}),~\mathrm{and}~(\ref{Problem:given_beam_SDR:4}).\nonumber
\end{align}
Although problem (P3.3) is non-convex, it can be shown similarly as for problem (P2.1), that (P3.3) can be solved equivalently by solving the following feasibility problem (P3.4) together with a bisection search over $t$.
\begin{align}
&\mathtt{(P3.4):}~\mathop\mathtt{find}~ \mv V \nonumber \\
&~~~~~~~~~~~~{\mathtt{s.t.}}
~\mathrm{Tr}({\mv R_{b,k,b,k}}\mv V) + |d_{b,k,b,k}|^2\ge\nonumber\\
&~~~~~~~~\alpha_{b,k} t\left[\sum\limits_{u\in\mathcal{K}_b, u\neq k}\mathrm{Tr}({\mv R_{b,u,b,k}}\mv V) + |d_{b,u,b,k}|^2 +\sum\limits_{i\in\mathcal B, i\neq b}\sum\limits_{u\in\mathcal{K}_i}\mathrm{Tr}({\mv R_{i,u,b,k}}V) + |d_{i,u,b,k}|^2 + \sigma^2_{b,k}\right],\nonumber\\
&~~~~~~~~~~~~~~~~~~ \forall k\in\mathcal{K}_b, b\in\mathcal B \label{Problem:given_beam_f:1}\\
&~~~~~~~~~~~~~~~~~~(\ref{Problem:given_beam_SDR:2}), (\ref{Problem:given_beam_SDR:3}),~\mathrm{and}~(\ref{Problem:given_beam_SDR:4}).\nonumber
\end{align}
Notice that problem (P3.4) is a convex semi-definite program (SDP) and thus can be solved optimally by using CVX \cite{CVX}. As a result, we have obtained the optimal solution to problem (P3.3), denoted by $\mv V^\star$ and $t^\star$.

Now, it remains to reconstruct the solution to problem (P3.2) or equivalently (P3.1)/(P3) based on $\mv V^\star$ and $t^\star$. In particular, if rank$(\mv V^\star)\le 1$, then $\mv V^\star$ and $t^\star$ are also the optimal solution to problem (P3.2). In this case, we have $\mv V^\star=\bar{\mv v}^\star \bar{\mv v}^{\star H}$, where $\bar{\mv v}^\star$ becomes the optimal solution to problem (P3.1). However, if rank$(\mv V^\star)>1$, then the following Gaussian randomization procedure \cite{Random} needs to be further adopted to produce a high-quality rank-one solution to problem (P3.2) or (P3.1). Specifically, suppose that the eigenvalue decomposition of $\mv V^\star$ is $\mv V^\star=\mv U\mv \Sigma\mv U^H$. Then, we set $\tilde{\mv v} = \mv U\mv\Sigma^{\frac{1}{2}}\mv r$, where $\mv r$ corresponds to a CSCG random vector with zero mean and covariance matrix $\mv I$, i.e., $\mv r\sim\mathcal{CN}(0,\mv I)$. Accordingly, we construct a feasible solution $\bar{\mv v}$ to problem (P3.1) as $[\bar{\mv{v}}]_n= e^{j\mathrm{arg}([\tilde{\mv{v}}]_n/[\tilde{\mv{v}}]_{N+1})}, \forall n\in\mathcal{N}$, where $[\bar{\mv{v}}]_n$ and $[\tilde{\mv{v}}]_n$ denote the $n$-th element of vector $\bar{\mv v}$ and $\tilde{\mv v}$, respectively. To guarantee the performance, the randomization process needs to be implemented over a large number of times and the best solution among them is selected as the obtained solution to problem (P3.1), denoted by ${\bar{\mv v}}^\star$. In this case, the obtained solution to problem (P3.2) is ${\bar{\mv v}}^\star {\bar{\mv v}}^{\star H}$. Based on the solution of ${\bar{\mv v}}^\star$ to problem (P3.1), we can accordingly obtain the solution of (P3) as $\mv v^\star$ based on (\ref{R_v}). Therefore, the SDR-based algorithm for solving problem (P3) is complete.

By alternately implementing the SDR-based solution to (P3) and the SOCP-based solution to (P2), we can obtain an efficient solution to the original problem (P1). We refer to this algorithm as the exact-alternating-optimization approach, which is summarized as Algorithm 1.

\begin{remark}
	It is worth noticing that the performance of the exact-alternating-optimization approach critically depends on the performance of the Gaussian randomization for SDR (when solving problem (P3)), especially when the rank of the obtained $\mv V^\star$ to SDP (P3.3) is larger than one. As such, the exact-alternating-optimization approach may lead to compromised performance, as the alternating optimization may terminate if the min-weighted SINR value decreases during iteration (due to the uncertainty in randomizations for SDR). Furthermore, the exact-alternating-optimization approach requires us to {\it exactly} solve the transmit and reflective beamforming subproblems (P2) and (P3) via solving a series of feasibility problems (P2.2) and (P3.4). Therefore, this approach also leads to very high computational complexity. These two drawbacks motive us to further develop an alternative approach with performance guarantee and lower computational complexity.
\end{remark}

\begin{table}[htp]
	\begin{center}
		\hrule
		\vspace{0.1cm}\textbf{Algorithm 1}: Exact-alternating-optimization approach for solving (P1) \vspace{0.1cm}
		\hrule \vspace{0.1cm}
		\begin{itemize}
			\item[1:] Initialize: $l=0$, $\mv v^{(0)}$ and accuracy threshold $\epsilon > 0$.
			\item[2:] {$\mv{\mathrm{Repeat}}$:}
			\item[3:] ~~~$l=l+1$;
			\item[4:] ~~~Under given $\mv v^{(l-1)}$, solve problem (P2) to obtain $\{\mv w_{b,k}^\star\}$ by solving a series of feasibility SOCP problems in (P2.2) together with a bisection search over $t$. Set $\mv w_{b,k}^{(l)}=\mv w_{b,k}^\star, \forall k\in\mathcal{K}_b, b\in\mathcal B$;
			\item[5:] ~~~Under given $\{\mv w_{b,k}^{(l)}\}$, solve problem (P3) to obtain $\mv v^\star$ by solving a series of feasibility SDP in (P3.4) together with a bisection over $t$, and adopting the randomization procedure. Set  $\mv v^{(l)}=\mv v^\star$;
			\item[6:] $\mv{\mathrm{Until}}$ the increase of the objective function in (P1) is smaller than  $\epsilon$ or the min-weighted-SINR value decreases.
		\end{itemize}
		\vspace{0.1cm} \hrule \label{Table:1}
	\end{center}
\end{table}

\section{Inexact Alternating Optimization}\label{sec:IV}
In this section, we propose an alternative design, namely the inexact-alternating-optimization approach, for solving the min-weighted-SINR maximization problem (P1) by overcoming the above drawbacks. Different from the above exact-alternating-optimization approach that alternately solves problems (P2) and (P3) exactly, in this alternative approach we only need to find an updated $\{\mv w_{b,k}\}$ and $\mv v$ to increase the min-weighted-SINR value at each iteration. In other words, suppose that at each particular iteration $l\ge 1$, the local point of $\{\mv w_{b,k}\}$ and $\mv v$ are denoted by $\{\mv w_{b,k}^{(l-1)}\}$ and ${\mv v}^{(l-1)}$, which correspond to the obtained $\{\mv w_{b,k}\}$ and $\mv v$ in the previous iteration. Then we aim to find $\{\mv w_{b,k}^{(l)}\}$ and $\mv v^{(l)}$ alternately at each iteration such that 
\begin{align}
\min_{k\in\mathcal{K}_b, b\in\mathcal B}\gamma_{b,k}(\{\mv w_{i,u}^{(l)}\}, \mv v^{(l)})/\alpha_{b,k}&\ge \min_{k\in\mathcal{K}_b, b\in\mathcal B}\gamma_{b,k}(\{\mv w_{i,u}^{(l-1)}\}, \mv v^{(l)})/\alpha_{b,k}\nonumber\\
&\ge \min_{k\in\mathcal{K}_b, b\in\mathcal B}\gamma_{b,k}(\{\mv w_{i,u}^{(l-1)}\}, \mv v^{(l-1)})/\alpha_{b,k}
\end{align}

\subsection{Inexact Coordinated Transmit Beamforming Update}
First, we update the coordinate transmit beamforming vectors $\{\mv w_{b,k}\}$. Inspired by \cite{Co_beam_2013}, instead of obtaining the exact optimal solution to problem (P2), we only need to find $\{\mv w_{b,k}^{(l)}\}$ with increased min-weighted-SINR value. In particular, under given $\mv{v}^{(l-1)}$ and
\begin{align}
t^{(l-1)}\!=\!\min_{k\in\mathcal{K}_b, b\in\mathcal B}\gamma_{b,k}(\{\mv w_{i,u}^{(l-1)}\}, \mv v^{(l-1)})/\alpha_{b,k},
\end{align}
we update the coordinated transmit beamforming vectors $\{\mv w_{b,k}^{(l)}\}$ as the optimal solution to the following problem
\begin{align}
\mathrm{(P4):}&~\mathop\mathtt{max}_{\{\mv w_{b,k}\},\xi} ~ \xi \nonumber \\
&~~~{\mathtt{s.t.}}
~~\frac{\mv a^H_{b,b,k}\mv w_{b,k}-\xi}{\sqrt{\sum\limits_{u\in\mathcal{K}_b, u\neq k}|\mv a^H_{b,b,k}\mv w_{b,u}|^2 +\sum\limits_{i\in\mathcal B, i\neq b}\sum\limits_{u\in\mathcal{K}_i}|\mv a^H_{i,b,k}\mv w_{i,u}|^2+\sigma_{b,k}^2}}\nonumber \\
&~~~~~~~~~\ge\sqrt{\alpha_{b,k} t^{(l-1)}}, \forall k\in\mathcal{K}_b, b\in\mathcal B \label{Problem:update_beam:1} \\
&~~~~~~~~~ \sum_{k\in\mathcal{K}_b}\|\mv w_{b,k}\|^2 \le P_b, \forall b\in\mathcal B, \label{Problem:update_beam:2}
\end{align}
where $\mv a_{i,b,k}=\mv\Phi_{i,b,k}^H\mv v^{(l-1)} + \mv h_{i,b,k}$ and $\xi$ is an auxiliary variable. Without loss of optimality, we choose the solution of $\{\mv w_{b,k}\}$ such that $\mv a^H_{b,b,k}\mv w_{b,k}$ becomes a non-negative value for any user $k\in\mathcal K_b$ in cell $b\in\mathcal{B}$. In this case, problem (P4) can be transformed as an SOCP similarly as for (P2.2), which is omitted here for brevity. Therefore, the optimal solution to (P4) can be obtained as $\{\mv w_{b,k}^\star\}$ and $\xi^\star$. 

Note that problem (P4) is always feasible, as $\mv{w}_{b,k}=\mv{w}_{b,k}^{(l-1)}, \forall k\in\mathcal{K}_b, b\in\mathcal B$, and $\xi=0$ correspond to one feasible solution. Therefore, at the optimal solution to (P4), we must have $\xi^\star \ge 0$. Therefore, by setting $\mv w_{b,k}^{(l)}=\mv{w}_{b,k}^\star, \forall k\in\mathcal{K}_b, b\in\mathcal B$, and combining them with ${\mv v}^{(l-1)}$, we have the achieved min-weighted-SINR as
\begin{align}
t^{\star} = \min_{k\in\mathcal{K}_b, b\in\mathcal B} \frac{\gamma_{b,k}( \{\mv w_{i,u}^{(l)}\}, {\mv v}^{(l-1)})}{\alpha_{b,k}}\ge t^{(l-1)}. \label{fm:update_sinr}
\end{align}
As a result, by solving problem (P4) once, we obtain an updated coordinated transmit beamforming with non-decreasing min-weighted-SINR value. As the original transmit beamforming optimization problem (P2) is not solved exactly (or optimally) in this case, we refer to this design as an inexact solution. As will be shown later, this design can not only reduce the computational complexity by avoiding the bisection search in the exact-alternating-optimization approach, but also lead to superior performance by jointly implementing the inexact reflective beamforming design next.

\subsection{Inexact Reflective Beamforming Update}
Next, we explain how to find an updated reflective beamforming vector $\mv v$ to increase the min-weighted-SINR without exactly solving problem (P3). This is implemented by applying the SCA technique. For notational convenience, we first define an auxiliary function for user $k\in\mathcal K_b$ in cell $b\in\mathcal{B}$ as
\begin{align}
&\mathcal{F}_{b,k}(\mv v, \{\mv w_{i,u}\}, t) \nonumber\\
&=\alpha_{b,k} t \left[\sum\limits_{u\in\mathcal{K}_b, u\neq k}\mv v^H\mv C_{b,u,b,k}\mv v + 2\mathrm{Re}\{\mv v^H\mv u_{b,u,b,k}\} + |d_{b,u,b,k}|^2\nonumber\right.\\
&\left.~~~~+\sum\limits_{i\in\mathcal B, i\neq b}\sum\limits_{u\in\mathcal{K}_i}\left(\mv v^H\mv C_{i,u,b,k}\mv v + 2\mathrm{Re}\{\mv v^H\mv u_{i,u,b,k}\} + |d_{i,u,b,k}|^2\right)+\sigma^2_{b,k}\right] \nonumber \\
&~~~~- \left(\mv v^H\mv C_{b,k,b,k}\mv v + 2\mathrm{Re}\{\mv v^H\mv u_{b,k,b,k}\} + |d_{b,k,b,k}|^2\right),\label{function}
\end{align}
where $\mv C_{i,u,b,k}$, $\mv u_{i,u,b,k}$, and $d_{i,u,b,k}$, $\forall k\in\mathcal{K}_b, u\in\mathcal{K}_i, b, i\in\mathcal B$ are defined in Section III-B. Note that at the local point $\{\mv w_{b,k}^{(l)}\}$ and $\mv v^{(l-1)}$, it must hold that $\min_{k\in\mathcal{K}_b, b\in\mathcal B} \mathcal F_{b,k}(\mv v^{(l-1)},\{\mv w_{i,u}^{(l)}\}, t^{\star}) = 0$. Accordingly, we update the reflective beamforming vector $\mv v$ at the IRS by solving the following problem:
\begin{align}
\mathtt{(P5):}&\mathop\mathtt{min}_{\mv v} ~\mathop\mathtt{max}_{k\in\mathcal{K}_b, b\in\mathcal B} ~\mathcal{F}_{b,k}\left(\mv v, \{\mv w_{i,u}^{(l)}\}, t^{\star}\right) \label{Problem:given_beam2:1} \\
&~{\mathtt{s.t.}}~~~|v_n|\le 1, \forall n\in\mathcal N. \label{Problem:given_beam2:2}
\end{align}
Notice that $\mv v^{(l-1)}$ is a feasible solution to problem (P5) with the achieved objective value being zero. Therefore, it is clear that the optimal solution to problem (P5) should be non-positive. 

However, problem (P5) is still non-convex as the objective function is non-convex with respect to $\mv v$. To address this issue, we apply the SCA technique to approximate the second convex term in the right-hand-side of (\ref{function}) by its first-order Taylor expansion. Note that a convex function is lower bounded by its first-order Taylor expansion at any given point. Therefore, at the given local point of $\mv v^{(l-1)}$, we have
\begin{align}
&\mathcal{F}_{b,k}\left(\mv v, \{\mv w_{i,u}^{(l)}\}, t^{\star}\right)\le \nonumber \\
&~~~\alpha_{b,k} t^{\star}\left[\sum\limits_{u\in\mathcal{K}_b, u\neq k}\mv v^H\mv C_{b,u,b,k}\mv v + 2\mathrm{Re}\{\mv v^H\mv u_{b,u,b,k}\} + |d_{b,u,b,k}|^2\nonumber\right.\\
&\left.~~~~+\sum\limits_{i\in\mathcal B, i\neq b}\sum\limits_{u\in\mathcal{K}_i}\left(\mv v^H\mv C_{i,u,b,k}\mv v + 2\mathrm{Re}\{\mv v^H\mv u_{i,u,b,k}\} + |d_{i,u,b,k}|^2\right)+\sigma^2_{b,k}\right] \nonumber \\
&~~~- \left({\mv v^{(l-1)}}^H\mv C_{b,k,b,k}{\mv v^{(l-1)}} + 2\mathrm{Re}\{{\mv v^{(l-1)}}^H\mv u_{b,k,b,k}\} + |d_{b,k,b,k}|^2\right) \nonumber \\
&~~~- 2\left(\mv C^H_{b,k,b,k}{\mv v^{(l-1)}} + \mv u_{b,k,b,k}\right)^H\left(\mv v - \mv v^{(l-1)}\right)\nonumber \\
&~~~\triangleq\mathcal{F}^{\mathtt{up}}_{b,k}\left(\mv v, \{\mv w_{i,u}^{(l)}\}, t^{\star},\mv v^{(l-1)}\right).\label{upper_bound}
\end{align}
By introducing an auxiliary variable $z$ and replacing $\mathcal{F}_{b,k}(\mv v, \{\mv w_{i,u}^{(l)}\}, t^{\star})$ by $\mathcal{F}^{\mathtt{up}}_{b,k}(\mv v, \{\mv w_{i,u}^{(l)}\}, t^{\star},\mv v^{(l-1)})$, problem (P5) is approximated as the following problem:
\begin{align}
\mathtt{(P5.1):}\mathop\mathtt{min}_{\mv v, z} ~& z \nonumber \\
{\mathtt{s.t.}}
~&\mathcal{F}^{\mathtt{up}}_{b,k}\left(\mv v, \{\mv w_{i,u}^{(l)}\}, t^{\star},\mv v^{(l-1)}\right)\le z, \forall i\in\mathcal K \label{Problem:given_beam_SCA:1} \\
~&|v_n|\le 1, \forall n\in\mathcal N. \label{Problem:given_beam_SCA:2}
\end{align}
Problem (P5.1) is a convex problem that can be solved optimally by CVX \cite{CVX}. Suppose that the optimal solution to problem (P5.1) is obtained as $\mv v^{\star\star}$ and $z^{\star\star}$. Notice that $\mv v^{(l-1)}$ and $z=0$ correspond to a feasible solution to (P5.1). Therefore, it is clear that we must have $z^{\star\star} \leq 0$, and equivalently $\min_{k\in\mathcal{K}_b, b\in\mathcal B}\mathcal{F}^\mathtt{up}_{b,k}(\mv v^{\star\star}, \{\mv w_{i,u}^{(l)}\}, t^{\star},\mv v^{(l-1)}) \le 0$. By using this together with (\ref{upper_bound}), we must have $\min_{k\in\mathcal{K}_b, b\in\mathcal B}\mathcal{F}_{b,k}(\mv v^{\star\star}, \{\mv w_{i,u}^{(l)}\}, t^{\star}) \le 0$. It thus follows that $\min_{k\in\mathcal{K}_b, b\in\mathcal B}\gamma_{b,k}(\mv v^{\star \star}, \{\mv w_{i,u}^{(l)}\})/\alpha_{b,k} \ge \min_{k\in\mathcal{K}_b, b\in\mathcal B} \gamma_{b,k}(\mv v^{(l-1)}, \{\mv w_{i,u}^{(l)}\})/\alpha_{b,k}$, i.e., $\mv v^{\star\star}$ leads to a non-decreasing min-weighted-SINR value. Therefore, we can directly update $\mv v$ as $\mv v^{\star\star}$, i.e., $\mv v^{(l)} = \mv v^{\star\star}$. Under given $\{\mv w_{b,k}^{(l)}\}$ together with ${\mv v}^{(l)}$, we denote the achieved minimum weighted SINR at users as $t^{(l)} = \min_{k\in\mathcal{K}_b, b\in\mathcal B} \gamma_{b,k}({\mv v}^{(l)}, \{\mv w_{i,u}^{(l)}\})/\alpha_{b,k}$. In summary, the inexact-alternating-optimization approach is presented as Algorithm 2.

\begin{table}[htp]
\begin{center}
\hrule
\vspace{0.1cm} \textbf{Algorithm 2}: Inexact-alternating-optimization approach for solving (P1)  \vspace{0.1cm}
\hrule \vspace{0.1cm}
\begin{itemize}
\item[1:] Initialize: $l=0$, $\mv v^{(0)}$, $t^{(0)}$ and accuracy threshold $\epsilon > 0$.
\item[2:] {$\mv{\mathrm{Repeat}}$:}
\item[3:] ~~~$l=l+1$;
\item[4:] ~~~Under given $\mv v^{(l-1)}$ and $t^{(l-1)}$, solve problem (P4) to obtain updated $\{\mv w_{b,k}^\star\}$ and $t^\star$, and set $\mv w_{b,k}^{(l)}=\mv w_{b,k}^\star, \forall k\in\mathcal{K}_b, b\in\mathcal B$;
\item[5:] ~~~Under given $\{\mv w_{b,k}^{(l)}\}$, $t^{\star}$, and $\mv v^{(l-1)}$, solve problem (P5.1) to obtain $\mv v^{\star\star}$, set $\mv v^{(l)}=\mv v^{\star\star}$ and update $t^{(l)}$;
\item[6:] $\mv{\mathrm{Until}}$ the increase of the objective function in (P1) is smaller than  $\epsilon$.
\end{itemize}
\vspace{0.1cm} \hrule\label{Table:2}
\end{center}
\end{table}

It is worth noting that in the inexact-alternating-optimization approach, the min-weighted-SINR value is monotonically non-decreasing after updating $\{\mv w_{b,k}\}$ and $\mv v$ at each iteration. As the optimal objective value of problem (P1) is bounded from above, it is clear that this approach is ensured to converge for problem (P1). This shows the performance advantage of this approach over the exact-alternating-optimization approach in Section III. Furthermore, notice that as only two convex optimization problems (one for updating the coordinated transmit beamforming vectors and the other for updating the reflective beamforming vector) need to be solved at each iteration, the  inexact-alternating-optimization approach clearly has lower computational complexity than the exact-alternating-optimization approach, as will be shown in numerical results later. 

\section{Low-Complexity Inexact Alternating Optimization}
In this section, we propose a low-complexity inexact-alternating-optimization design to further reduce the computational complexity. In this design, the coordinated transmit beamforming vectors $\{\mv{w}_{b,k}\}$ are updated inexactly based on that in Section IV-A. Therefore, we only need to focus on the update of the reflective beamforming vector $\mv v$ by solving problem (P3). 

Notice that in the inexact-alternating-optimization in Section IV-B, for updating the reflective beamforming vector, we first reformulate problem (P3) into an equivalent problem (P5), and then we solve problem (P5) inexactly by solving problem (P5.1) based on the principle of SCA. However, problem (P5.1) is solved exactly with an optimal solution by using CVX, which may lead to relatively high computational complexity due to the interior-point method implemented in CVX. To further reduce the complexity, in the following, we adopt the subgradient projection method \cite{subgradient} to solve problem (P5.1) inexactly.

To facilitate the design, we reformulate problem (P5.1) as 
\begin{align}
\mathtt{(P5.2):}\mathop\mathtt{min}_{\mv v} ~& \mathop\mathtt{max}_{k\in\mathcal{K}_b, b\in\mathcal B}~\mathcal{F}^{\mathtt{up}}_{b,k}\left(\mv v, \{\mv w_{i,u}^{(l)}\}, t^{\star},\mv v^{(l-1)}\right) \nonumber \\
{\mathtt{s.t.}}
~&~~~|v_n|\le 1, \forall n\in\mathcal N. \label{Problem:SPM:1}
\end{align}
Notice that problem (P5.2) is a constrained convex optimization problem, where $\mathcal{C}=\{\mv v~|~|v_n|\le 1, \forall n\in\mathcal N\}$ is a convex set. Let $\mathcal{G}(\mv v)$ denote the objective function of problem (P5.2), i.e., $\mathcal{G}(\mv v) = \max_{k\in\mathcal{K}_b, b\in\mathcal B}~\mathcal{F}^{\mathtt{up}}_{b,k}(\mv v, \{\mv w_{i,u}^{(l)}\}, t^{\star},\mv v^{(l-1)})$, where $\mathcal{F}^{\mathtt{up}}_{b,k}(\mv v, \{\mv w_{i,u}^{(l)}\}, t^{\star},\mv v^{(l-1)}), \forall k\in\mathcal{K}_b, b\in\mathcal B$ are all differentiable quadratic functions. Let $\partial{\mathcal G}({\mv{v}})$ denote the subdifferential of $\mathcal{G}({\mv{v}})$ at point $\mv{v}$. Then we have
\begin{align}
\partial\mathcal{G}(\mv v)=\mathrm{\mv{Conv}}\left\{\nabla\mathcal{F}^{\mathtt{up}}_{b,k}(\mv v, \{\mv w_{i,u}^{(l)}\}, t^{\star},\mv v^{(l-1)})~|~\mathcal{F}^{\mathtt{up}}_{b,k}(\mv v, \{\mv w_{i,u}^{(l)}\}, t^{\star},\mv v^{(l-1)})=\mathcal{G}(\mv v)\right\}, \label{convex_hull}
\end{align}
i.e., the subdifferential of the maximum of functions is the convex hull of the union of gradients of the ‘active’ functions at the point $\mv{v}$ \cite{subgradient}. Let $\mv{g}(\mv{v})$ denote any subgradient of $\mathcal{G}({\mv{v}})$ at the point $\mv{v}$, i.e., $\mv{g}(\mv{v})\in\partial\mathcal{G}(\mv v)$. Without loss of generality, one subgradient of the objective function $\mathcal{G}({\mv{v}})$ can be chosen as the gradient of one of the functions that achieves the maximum at the point ${\mv{v}}$, i.e., 
\begin{align}
\mv{g}(\mv{v})&=\nabla\mathcal{F}^{\mathtt{up}}_{b,k}(\mv v, \{\mv w_{i,u}^{(l)}\}, t^{\star},\mv v^{(l-1)})\nonumber\\
&=\alpha_{b,k} t^{\star}\left[\sum\limits_{u\in\mathcal{K}_b, u\neq k}\left(2\mv C_{b,u,b,k}^H{\mv{v}} +  2\mv u_{b,u,b,k}\right)+\sum\limits_{i\in\mathcal B, i\neq b}\sum\limits_{u\in\mathcal{K}_i}\left(2\mv C_{i,u,b,k}^H{\mv{v}} +  2\mv u_{i,u,b,k}\right)\right]\nonumber\\
&~~~~ - 2\left(\mv C^H_{b,k,b,k}{\mv v^{(l-1)}} + \mv u_{b,k,b,k}\right), \label{subgradient}
\end{align}
where $\mathcal{F}^{\mathtt{up}}_{b,k}({\mv{v}}, \{\mv w_{i,u}^{(l)}\}, t^{\star},\mv v^{(l-1)}) = \mathcal{G}({\mv{v}})$. Let $\mathtt{Proj}_\mathcal{C}(\mv{x})$ denote the (Euclidean) projection onto the set $\mathcal{C}$ from a given point $\mv{x}=[x_1,...,x_N]^T$, i.e., 
$\mathtt{Proj}_\mathcal{C}(\mv{x})\triangleq\mathrm{arg}\mathop\mathtt{min}_{\mv{v}\in\mathcal{C}} \|\mv{v}-\mv{x}\|$. Since the reflection coefficients of all units are fully separable in constraint (\ref{Problem:SPM:1}), each element of $\mathtt{Proj}_\mathcal{C}(\mv{x})$ is given by
\begin{align}
\left[\mathtt{Proj}_\mathcal{C}(\mv{x})\right]_n =
\begin{cases} 
\frac{x_n}{|x_n|},  & \mbox{if }|x_n|>1,\\
x_n, & \mbox{otherwise.}
\end{cases} \label{projection}
\end{align}
For notational convenience, suppose that at each iteration $k\ge0$ of the subgradient projection method, $\hat{\mv{v}}^{(k)}$ is the obtained reflective beamforming vector given by
\begin{align}
\hat{\mv{v}}^{(k)} = \mathtt{Proj}_\mathcal{C}(\hat{\mv{v}}^{(k-1)}-\lambda_k\mv{g}^{(k)}(\hat{\mv{v}}^{(k-1)})),\label{next_step}
\end{align}
where $\lambda_k>0$ is the step size at the $k$-th iteration. In particular, we use the constant step length rule\footnote{Generally, there are many other different types of step size rules in the subgradient projection algorithm, such as constant step size, diminishing step size, and others. However, via numerical results, it is shown that compared with the others, the constant step length rule achieves the best objective value more efficiently.}, i.e.,
\begin{align}
\lambda_k=\frac{\gamma}{\|\mv{g}^{(k)}\|_2}, \label{step_size}
\end{align}
where $\gamma$ is a constant. 

Since the subgradient projection method is generally not a desent method, we have to keep track of the best point $\hat{\mv{v}}_{\mathrm{best}}$ found so far. Let $\mathcal{G}(\hat{\mv{v}}_{\mathrm{best}})$ denotes the best objective function value found so far, i.e., 
\begin{align}
\mathcal{G}(\hat{\mv{v}}_{\mathrm{best}})=\min\{\mathcal{G}(\hat{\mv{v}}^{(1)}),...,\mathcal{G}(\hat{\mv{v}}^{(k)})\}. \label{best_obj_value}
\end{align} 
Therefore, we have $\hat{\mv{v}}_{\mathrm{best}} = \hat{\mv{v}}$ if $\mathcal{G}(\hat{\mv{v}})=\mathcal{G}(\hat{\mv{v}}_{\mathrm{best}})$, and it is also the solution to problem (P5.2), i.e., $\mv{v}^{(l)}=\hat{\mv{v}}_{\mathrm{best}}$. In order to reduce the computational complexity, the subgradient projection method only needs to be performed over a few iterations to update the reflective beamforming vector $\mv{v}$. As this design inexactly solves problem
(P5.1), it is expected to be significantly cheaper than exactly
solving problem (P5.1) in the inexact-alternating-optimization approach. In summary, the low-complexity inexact-alternating-optimization approach is presented as Algorithm 3.

Now, we compare the computational complexity of updating the reflective beamforming vector $\mv{v}$ in Algorithm 3 in Table III, versus that in Algorithm 2 in  Table II. In Algorithm 3, the time complexity of solving problem (P5.1) is dominated by the subgradient projection method in steps 6-9. Specifically, the complexity of step 8 is of order $KN$ (because all matrices $\mv C_{i,u,b,k}$ are of rank one), i.e., $\mathcal{O}(KN)$, and note that step 9 iterates $T$ times to terminate. Therefore, the complexity of the updating reflective beamforming vector at each iteration $l$ in Algorithm 3 is $\mathcal{O}(KNT)$. However, in Algorithm 2 of the inexact-alternating-optimization approach, we use the interior-point algorithm (implemented in CVX) to solve problem (P5.1), for which the time complexity of updating $\mv{v}$ is $\mathcal{O}\left((K+N)^{\frac{1}{2}}N(N^2+K^3)\right)$ \cite[page 423]{complexity}. Therefore, the subgradient projection method has a lower computational complexity. As a result, the low-complexity inexact-alternating-optimization approach in this section achieves lower computational complexity than the inexact-alternating-optimization in Section IV.

\begin{table}[htbp]
	\begin{center}
		\hrule
		\vspace{0.1cm} \textbf{Algorithm 3}: Low-complexity inexact-alternating-optimization approach for solving (P1)  \vspace{0.1cm}
		\hrule \vspace{0.1cm}
		\begin{itemize}
			\item[1:] Initialize: $l=0$, $T$, $\mv v^{(0)}$, $t^{(0)}$ and accuracy threshold $\epsilon > 0$.
			\item[2:] {$\mv{\mathrm{Repeat}}$:}
			\item[3:] ~~~$l=l+1$;
			\item[4:] ~~~Under given $\mv v^{(l-1)}$ and $t^{(l-1)}$, solve problem (P4) to obtain $\{\mv w_{b,k}^\star\}$ and $t^\star$. Set $\mv w_{b,k}^{(l)}=\mv w_{b,k}^\star, \forall k\in\mathcal{K}_b, b\in\mathcal B$;
			\item[5:] ~~~Initialize: $\hat{\mv{v}}^{(0)}=\mv v^{(l-1)},m=0$
			\item[6:] ~~~{$\mv{\mathrm{Repeat}}$:}
			\item[7:] ~~~~~~~$m=m+1$;
			\item[8:] ~~~~~~~Under given $\{\mv w_{b,k}^{(l)}\}$, $t^\star$, $\mv{v}^{(l-1)}$ and $\hat{\mv{v}}^{(m-1)}$, update $\hat{\mv{v}}^{(m)}$ based on (\ref{next_step});
			\item[9:] ~~~$\mv{\mathrm{Until}}$ $m\ge T$, obtain the best point $\hat{\mv{v}}_{\mathrm{best}}$ among all $\hat{\mv{v}}^{(m)}, m=1,\dots,T$, and set $\mv{v}^{(l)}=\hat{\mv{v}}_{\mathrm{best}}$.
			\item[10:] $\mv{\mathrm{Until}}$ the increase of the objective function in (P1) is smaller than  $\epsilon$.
		\end{itemize}
		\vspace{0.1cm} \hrule\label{Table:3}
	\end{center}
\end{table}

It is worth noting that for the algorithm of low-compelxity inexact-alternating-optimization, the min-weighted-SINR is
always non-decreasing after each update of $\{\mv w_{b,k}\}$ and $\mv v$. Therefore, the objective value of (P1) is ensured to be non-decreasing after each iteration. As a result, the low-complexity inexact-alternating-optimization approach is also ensured to converge for problem (P1).

\begin{figure}[htbp]
	\centering
	\epsfxsize=1\linewidth
	\includegraphics[width=9cm]{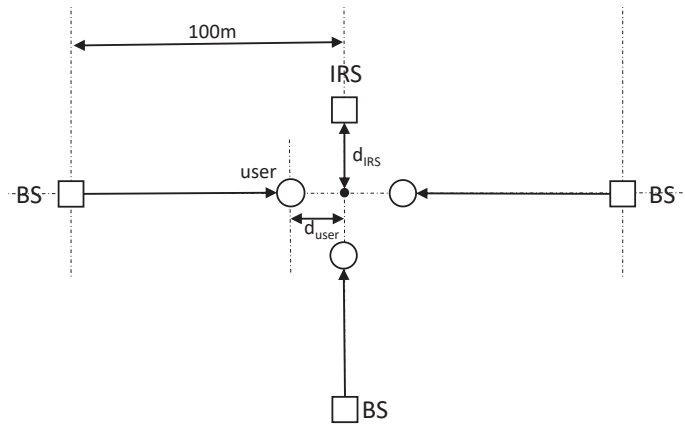}
	\caption{Simulation setup. } \label{fig:simulation_setup}
	\vspace{-1em}
\end{figure}

\section{Numerical Results}
In this section, we provide numerical results to evaluate the performance of the proposed approaches in the IRS-aided multi-cell MISO system. In the simulation, we first consider the scenario with $K_b = 1, \forall b\in \mathcal B$, i.e., there is one single user at each cell unless otherwise stated. In particular, we consider a scenario as shown in Fig. \ref{fig:simulation_setup}, where $B=3$ BSs are located at $(-100~\text{m},0)$, $(100~\text{m},0)$, and $(0,100~\text{m})$\footnote{Notice that with advancements of the ultra-dense deployment of small base stations (e.g., micro, pico, and femto BSs), the distances among BSs are normally 200 meters or less \cite{UDN}. In this case, our proposed design is meaningful in practice.}, respectively, each equipped with $M=3$ antennas. We consider symmetrically distributed users unless otherwise stated, where the three users are located at $(-d_\text{user},0)$, $(d_\text{user},0)$, and $(0,d_\text{user})$, with $d_\text{user} = 5~\text{m}$. An IRS with $N$ reflecting units is deployed at $(0,-d_\text{IRS})$, with $d_\text{IRS} = 10~\text{m}$. Furthermore, we set the maximum transmit power at all BSs to be identical, i.e., $P_b=P_\text{max}, \forall b\in\mathcal B$, and we are interested in the minimum SINR at users by setting $\alpha_{b,k}=1,\forall k\in\mathcal{K}_b, b\in\mathcal B$. In addition, we consider the distance-dependent path loss model as
\begin{align}
P_L=C_0\left(\frac{d}{d_0}\right)^{-\alpha}, \label{fm:16}
\end{align}
where $C_0=-30~$dB denotes the path loss at the reference distance of $d_0=1~$m, $\alpha$ denotes the path loss exponent, $d$ denotes the distance between the transmitter and the receiver. For the BS-user, BS-IRS, and IRS-user links, we set the path-loss exponents $\alpha$ to be $3.6$, $2$, and $2.5$, respectively. Furthermore, we consider Rayleigh fading for the BS-user and IRS-user links and the Rician fading for the BS-IRS link. Accordingly, we have
\begin{align}
\mv G_b=\sqrt{\frac{K_R}{1+K_R}}\mv{G}_b^{\text{LOS}}+\sqrt{\frac{1}{1+K_R}}\mv{G}_b^{\text{NLOS}},\forall b\in\mathcal{B},\label{fm:Ricain}
\end{align}
where $\mv{G}_b^{\text{LOS}}\in\mathbb{C}^{N\times M}$ is the LOS deterministic component, $\mv{G}_b^{\text{NLOS}}\in\mathbb{C}^{N\times M}$ is the non-LOS Rayleigh fading component, and $K_R\ge 0$ denotes the Rician factor. The noise power at each user $i$ is set as $\sigma^2_{b,k}=-80~{\text{dBm}}, \forall k\in\mathcal{K}_b, b\in\mathcal B$. The constant step length in (\ref{step_size}) is set as $\gamma=0.01$. The number of iterations for the subgradient projection method is set as $T=100$.  All the results are averaged over 100 independent channel realizations.

\begin{figure}[htbp]
	\centering
	\epsfxsize=1\linewidth
	\includegraphics[width=9cm]{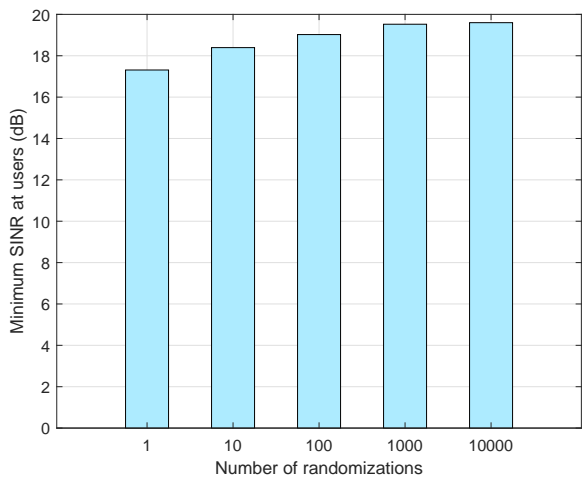}
	\caption{The minimum SINR at users versus the number of randomizations used for SDR in the exact-alternating-optimization approach. } \label{fig:Randomization}
	\vspace{-1em}
\end{figure}

First, we evaluate the effect of randomization for SDR in the exact-alternating-optimization approach. Fig. \ref{fig:Randomization} shows the minimum SINR at users versus the number of Gaussian randomizations. It is observed that the achieved min-SINR value by the exact-alternating-optimization approach increases as the number of randomizations increases. This shows the importance of using a large number of randomizations to mitigate the resultant uncertainty. To balance between the performance and complexity, we use 1000 Gaussian randomizations in this paper.

\begin{figure}[htbp]
	\centering
	\epsfxsize=1\linewidth
	\includegraphics[width=9cm]{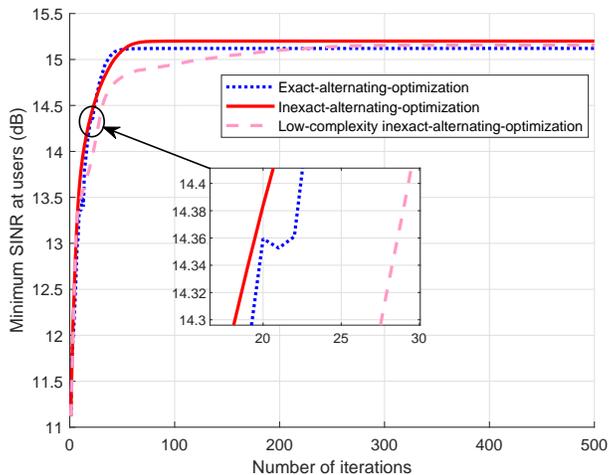}
	\caption{Convergence behavior of the three proposed alternating-optimization-based algorithms.} \label{fig:Iteration}
	\vspace{-1em}
\end{figure}

Fig. \ref{fig:Iteration} shows the convergence behaviour of our proposed three approaches, where the maximum transmit power is set as $P_\text{max}= 35~$dBm. It is observed that the inexact-alternating-optimization and the low-complexity inexact-alternating-optimization approaches give monotonically increasing min-SINR values over iterations, thus leading to a converged solution. By contrast, the exact-alternating-optimization approach terminates with a much lower min-SINR value than that achieved by the inexact-alternating-optimization approach. This is because that the SDR results in a reduced min-SINR value during the iteration due to the uncertainty in Gaussian randomization.  

Next, we compare our proposed designs with the following four benchmark schemes. As only a single user is considered in each cell in the comparison, we rewrite $\mv{w}_{i,u}$ and $\mv{a}_{i,b,k}$ as $\mv{w}_b$ and $\mv{a}_{i,b}$, respectively, for simplicity, by omitting the user index $k$ for each cell.
\begin{itemize}
	\item {\bf Inexact alternating optimization with MRT:} In this scheme, the coordinated transmit beamforming vectors at BSs are  set based on the MRT principle, i.e., $\mv{w}_b=\frac{\mv{a}_{b,b}}{\|\mv{a}_{b,b}^H\|},\forall b\in\mathcal{B}$; while the relective beamforming is updated by solving problem (P5.1). The transmit and reflective beamforming optimizations are implemented in an alternating manner until the increase of the objective function in problem (P1) is smaller than a certain threshold or the min-weighted-SINR value decreases.
	\item {\bf Inexact alternating optimization with ZF:} In this scheme, the coordinated transmit beamforming vectors at BSs are set based on the ZF principle, i.e., $|\mv{a}_{i,b}^H\mv{w}_i|=0,\forall i,b\in\mathcal{B}, i\neq b$. Accordingly, we have 
	\begin{align}
	\mv{w}_b=(\mv I-\mv{A}_b(\mv{A}_b^H\mv{A}_b)^{-1}\mv{A}_b^H)\mv{a}_{b,b}, \forall b\in\mathcal{B}, \label{fm:ZF_beam}
	\end{align}
	where $\mv{A}_b=[\mv{a}_{1,b},\dots,\mv{a}_{b-1,b},\mv{a}_{b+1,b},\dots,\mv{a}_{K,b}]$ \cite{ZF}. On the other hand, the relective beamforming vectors are obtained by solving problem (P5.1). The transmit and reflective beamforming optimizations are implemented in an alternating manner until convergence or the min-weighted-SINR value decreases.
	\item {\bf Benchmark scheme with random reflective beamforming:} In this scheme, we set the phase shift $\theta_n$ for each reflecting unit $n\in\mathcal{N}$ at the IRS as a uniformly distributed random value in $[0, 2\pi)$, and set $\beta_n = 1, \forall n\in\mathcal{N}$. Under such given reflective beamforming, we solve problem (P2) to obtain the corresponding coordinated transmit beamforming.
	\item {\bf Benchmark scheme without IRS}: Without IRS deployed, we only need to optimize the coordinated transmit beamforming vectors by solving problem (P2), in which $\{\mv a_{i,b,k}\}$ is replaced as $\{\mv h_{i,b,k}\}$.
\end{itemize}

\begin{figure}[htbp]
	\centering
	\epsfxsize=1\linewidth
	\includegraphics[width=9cm]{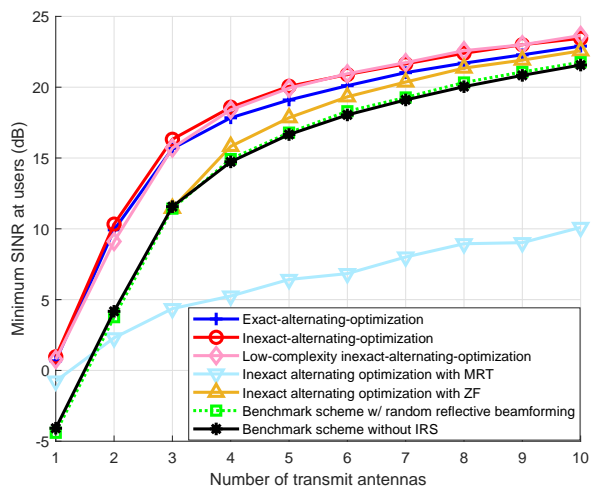}
	\caption{The minimum SINR at users versus the number of transmit antennas $M$ at each BS. } \label{fig:antenna_number}
	\vspace{-1em}
\end{figure}

Fig. \ref{fig:antenna_number} shows the minimum SINR at users versus the number of transmit antennas $M$ at each BS. First, it is observed that the proposed three approaches considerably outperform the four benchmark schemes. This shows the benefit of our proposed designs. Similarly as in Fig. \ref{fig:Iteration}, the inexact-alternating-optimization approach is observed to considerably outperform the exact-alternating-optimization one. It is also observed that the performance achieved by the exact-alternating-optimization-with-ZF scheme approaches the three proposed designs as $M$ becomes large. This is due to the fact that the ZF transmit beamforming becomes asymptotically optimal as the number of transmit antennas becomes large. Furthermore, it is observed that the benchmark scheme with random beamforming has a similar performance as that without IRS.
This shows that the benefit of IRS can only be achieved under proper reflective beamforming optimization.

\begin{figure}[htbp]
	\centering
	\subfigure{
		\begin{minipage}[t]{0.47\linewidth}
			\centering\setcounter{figure}{5} 
			\includegraphics[width=8cm]{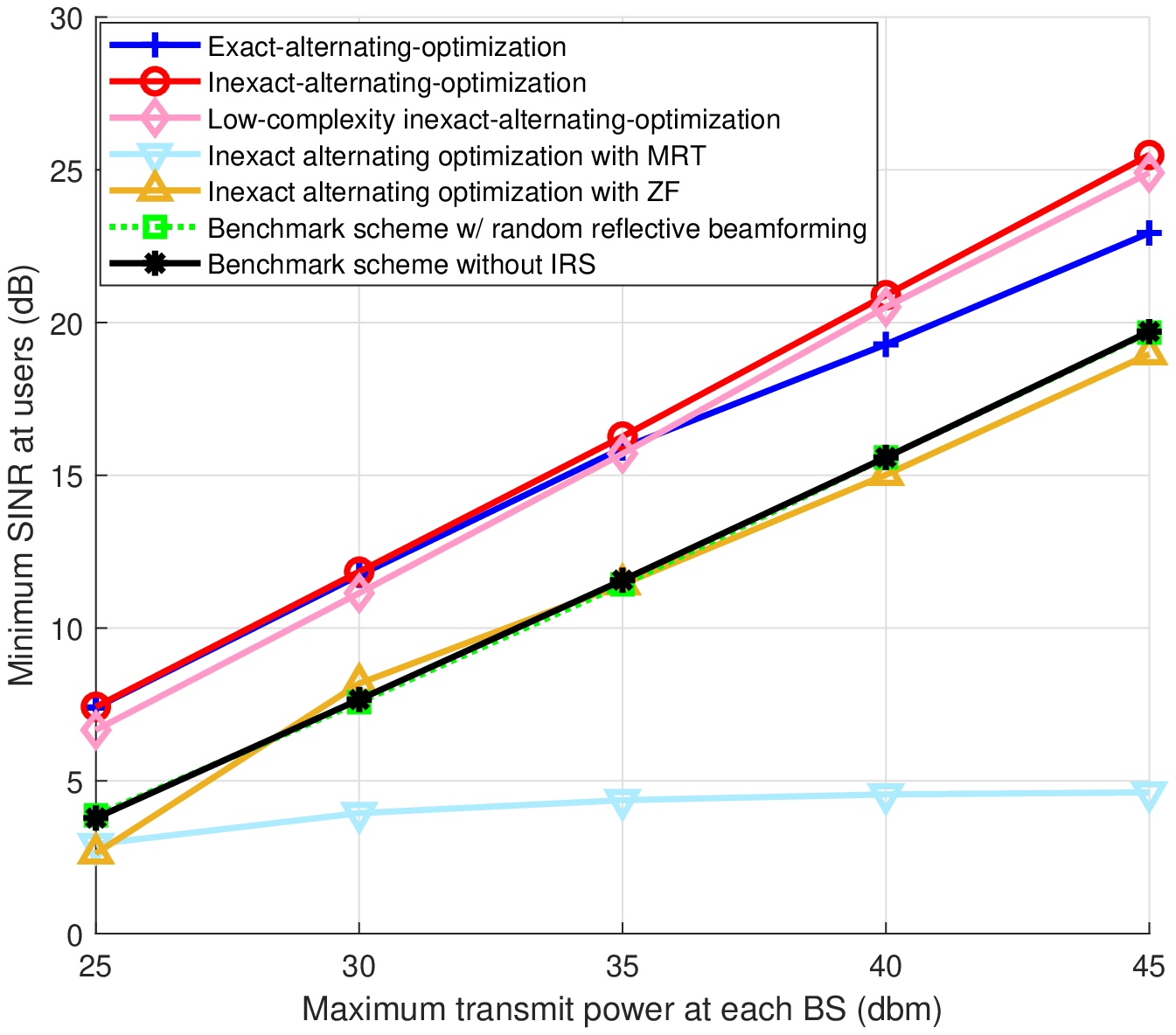}
			\caption{The minimum SINR at users versus the maximum transmit power $P_{\text{max}}$ at each BS.}\label{fig:Power}
		\end{minipage}
	}
	\subfigure{
		\begin{minipage}[t]{0.47\linewidth}
			\centering
			\includegraphics[width=8cm]{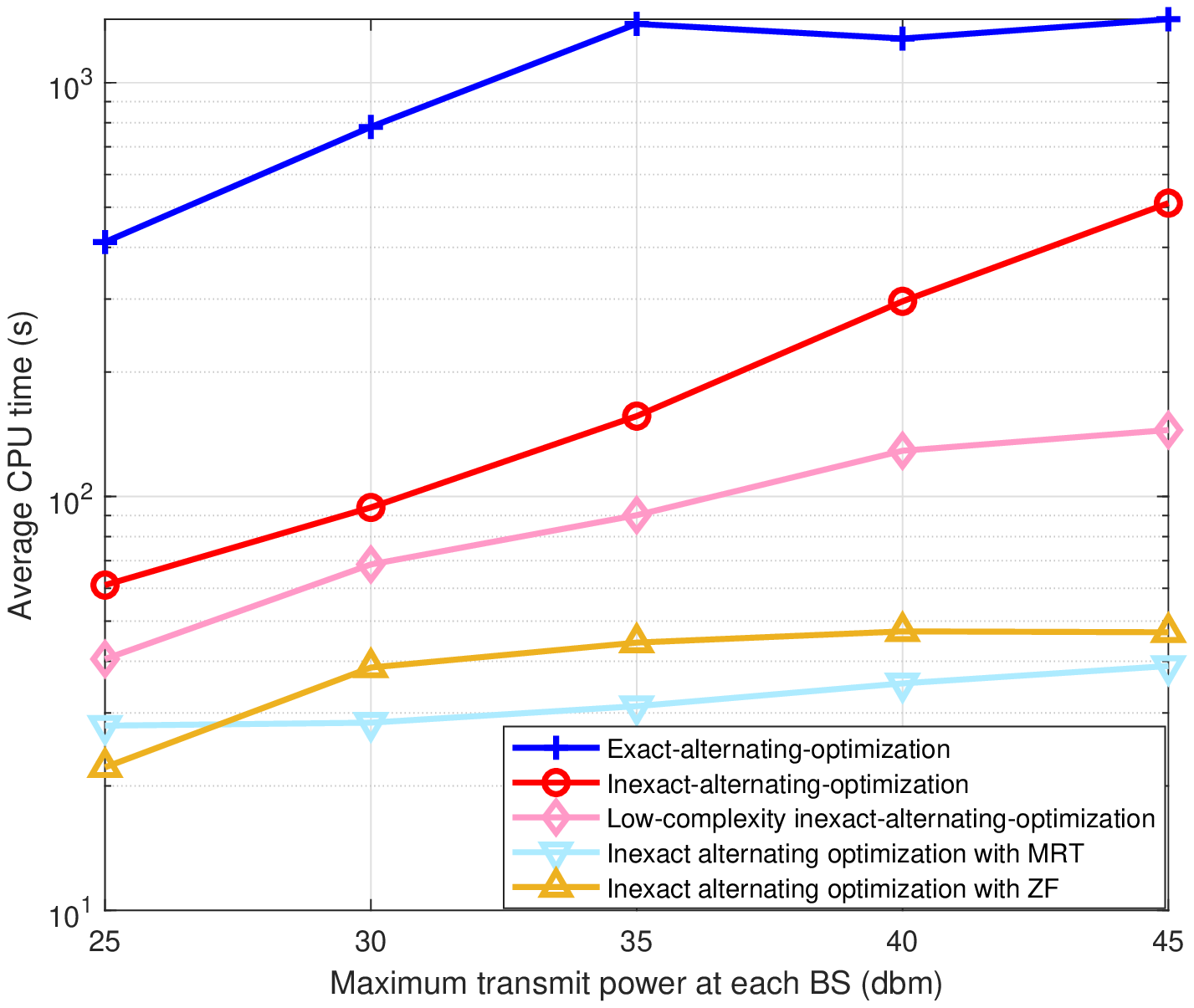}
			\caption{The average CPU time versus the maximum transmit power $P_{\text{max}}$ at each BS.}\label{fig:Power_Times}
		\end{minipage}
	}
	\vspace{-0.2cm}
	\centering
	\vspace{-0.5cm}
\end{figure}


Fig. \ref{fig:Power} shows the minimum SINR at users versus the maximum transmit power $P_\text{max}$ at each BS. Similar observations are made as in Fig. \ref{fig:antenna_number}. Specifically, the exact-alternating-optimization approach is observed to perform inferior to the inexact-alternating-optimization and low-complexity inexact-alternating-optimization approaches when $P_{\text{max}}$ becomes large (e.g., $P_{\text{max}}=45~$dBm), due to the uncertainty in Gaussian randomization. It is also observed that the ZF transmit beamforming is not asymptotically optimal any longer in the high SINR regime. This is due to the fact that the IRS can adjust its reflective beamforming to mitigate the co-channel interference, thus making the ZF transmit beamforming highly suboptimal in general, even in the high SINR regime.

Fig. \ref{fig:Power_Times} show the average CPU time versus the maximum transmit power $P_\text{max}$ at each BS. It is observed that the inexact-alternating-optimization approach takes much less time than the exact-alternating-optimization one, with superior performance at the same time. Furthermore, the low-complexity inexact-alternating-optimization design is observed to take even much less CPU time than the other two proposed approaches with slightly compromised performance, while it is also observed to have similar complexity as the two benchmark schemes of exact-alternating-optimization with ZF and MRT, but with much better performance. 

\begin{figure}[htbp]
	\centering
	\subfigure{
		\begin{minipage}[t]{0.47\linewidth}
			\centering\setcounter{figure}{7} 
			\includegraphics[width=8cm]{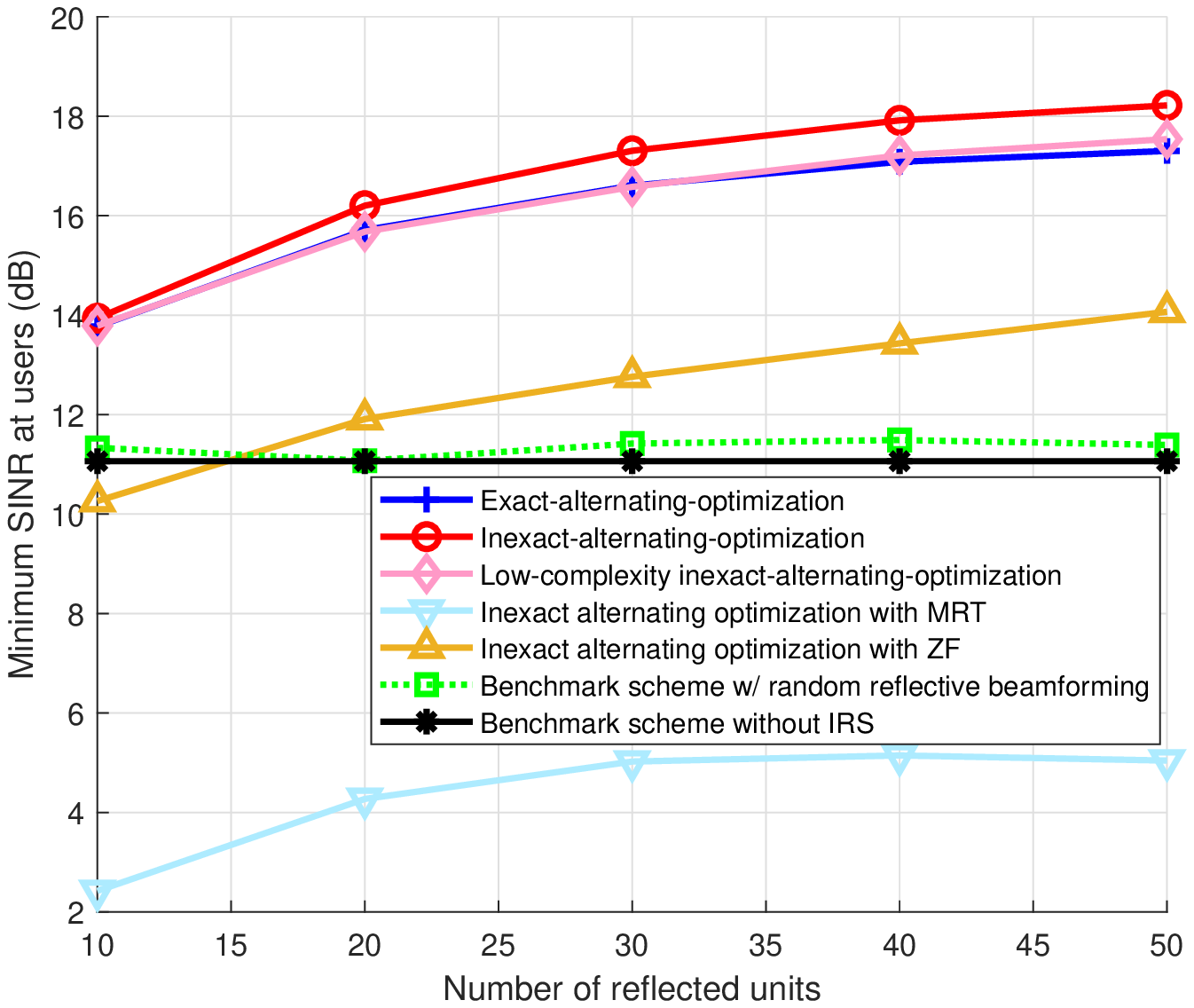}
			\caption{The minimum SINR at users versus the number of reflecting units $N$ at the IRS.}\label{fig:units}
		\end{minipage}
	}
	\subfigure{
		\begin{minipage}[t]{0.47\linewidth}
			\centering
			\includegraphics[width=8cm]{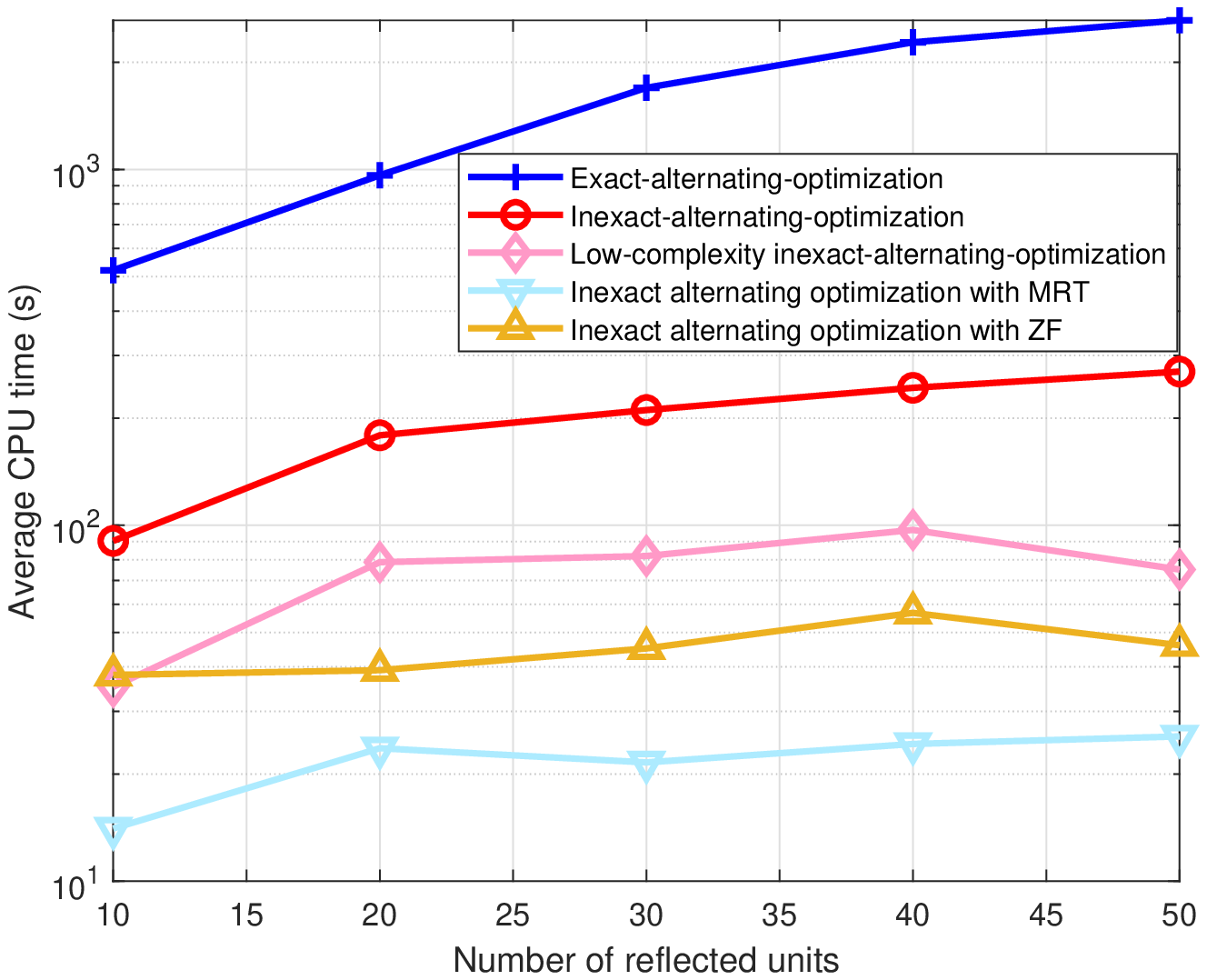}
			\caption{The average CPU time versus the number of reflecting units $N$ at the IRS.}\label{fig:units_time}
		\end{minipage}
	}
	\vspace{-0.2cm}
	\centering
	\vspace{-0.5cm}
\end{figure}

%

Fig. \ref{fig:units} shows the minimum SINR at users versus the number of reflecting units $N$ at the IRS, where we set $P_\text{max}=35$ dB. It is shown that the resulting min-SINR values by the three proposed approaches increase as $N$ becomes larger. It is also observed that the performance gap between the exact- and inexact-alternating-optimization approaches becomes larger as $N$ increases. Fig. \ref{fig:units_time} shows the average CPU time versus the number of reflecting units $N$ at the IRS. Similar observations can be made as in Fig. \ref{fig:Power_Times}. 

\begin{figure}[htbp]
	\centering\setcounter{figure}{9} 
	\epsfxsize=1\linewidth
	\includegraphics[width=9cm]{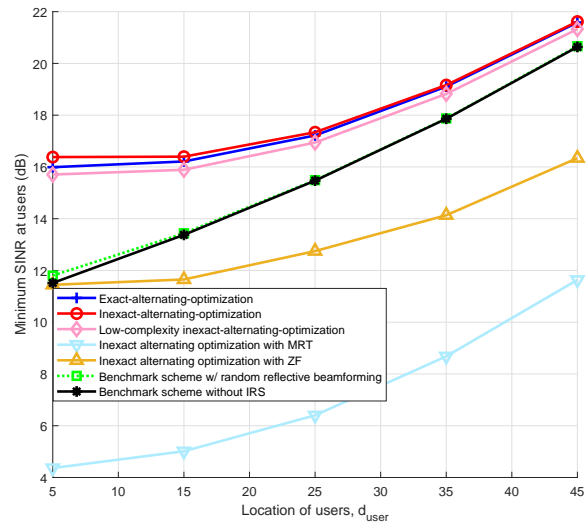}
	\caption{The minimum SINR at users versus the location of users $d_\text{user}$. } \label{fig:Distence}
	\vspace{-1em}
\end{figure}

Fig. \ref{fig:Distence} shows the minimum SINR at users versus the location of users $d_\text{user}$. It is observed that as $d_\text{user}$ increases (or equivalently, the users move towards the cell center),  the performance gains of the three proposed approaches over the benchmark schemes decrease. This is due to the fact that in this case, the direct communication link from each BS to the corresponding user becomes strong, and thus the gain brought by the IRS becomes less significant.

\begin{figure}[htbp]
\centering
 \epsfxsize=1\linewidth
    \includegraphics[width=9cm]{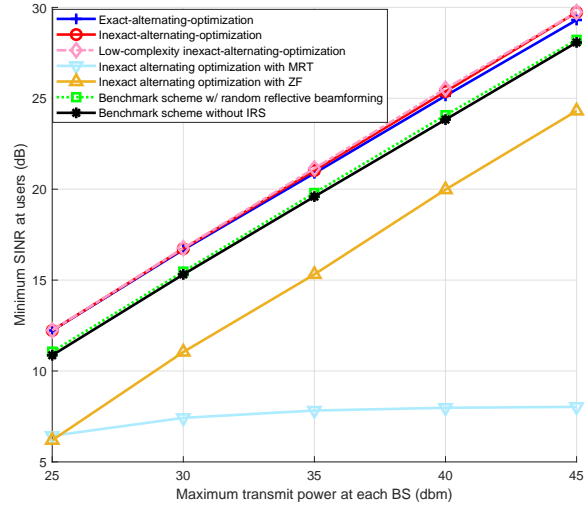}
\caption{The minimum SINR at users versus the maximum transmit power $P_{\text{max}}$ at each BS,  in the scenario with randomly distributed users. } \label{fig:Random}
\vspace{-1em}
\end{figure}

To further reveal the practical performance, Fig. \ref{fig:Random} shows the minimum SINR at users versus $P_\text{max}$, in the scenario where the three users are randomly distributed within a triangle area whose vertices correspond to the three BSs. It is observed that the IRS results in $68.4\%$ performance gains as compared to the benchmark scheme without the IRS when $P_\text{max}=35~$dBm. 

\begin{figure}[htbp]
	\centering
	\epsfxsize=1\linewidth
	\includegraphics[width=9cm]{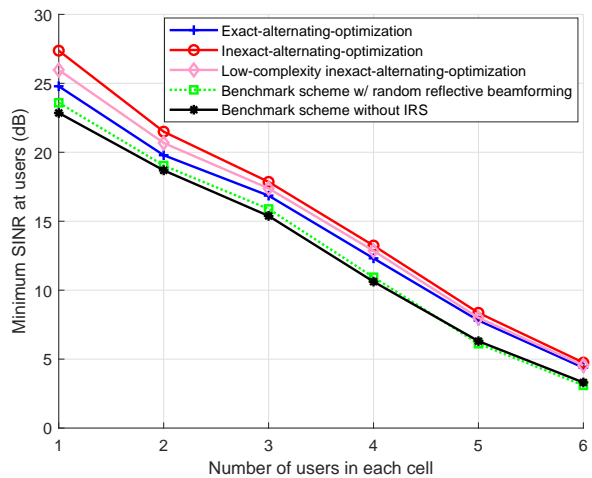}
	\caption{The minimum SINR at users versus the number of users in each cell. } \label{fig:Multiuser}
	\vspace{-1em}
\end{figure}

Next, we evaluate the performance of our proposed designs in the scenario with multiple users in each cell. We consider the case with two BSs located at $(-100~\text{m},0)$ and $(100~\text{m},0)$, respectively, in which the users in each cell are randomly distributed at the cell edge, within a circular area with radius being $10~\text{m}$ and centers $(-10~\text{m}, 0)$ and $(10~\text{m}, 0)$, respectively. We set $M = 8$, $N = 20$, and $P_{max}=35~$dBm. Under this setup, Fig. \ref{fig:Multiuser} shows the minimum SINR among all users versus the number of users $\bar{K}$ in each cell, where $K_b = \bar{K}, \forall b\in \mathcal B$. It is observed that the three proposed algorithms outperform the benchmark schemes, similarly as observed for the case with a single user at each cell.


\begin{figure}[htbp]
	\centering
	\epsfxsize=1\linewidth
	\includegraphics[width=9cm]{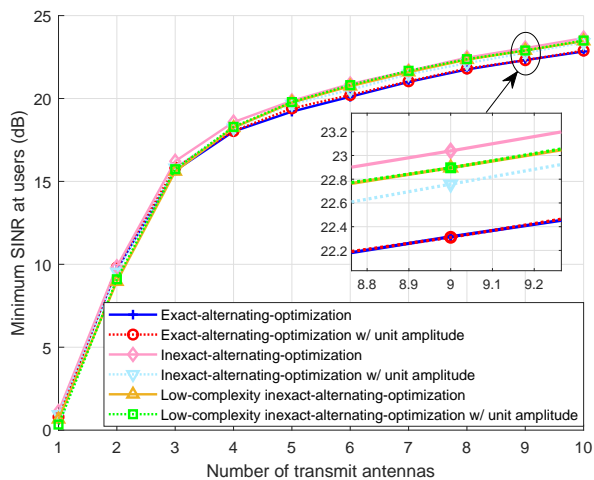}
	\caption{Performance comparison between the proposed algorithms with and without unit-amplitude reflection constraints.} \label{fig:antenna_number_unit_amplitude}
	\vspace{-1em}
\end{figure}

Finally, we show that our proposed designs with joint phase and amplitude control are also applicable for the case with only phase control (i.e., the reflecting amplitudes at reflecting units are $|v_n| = 1$). In particular, after obtaining the optimized reflective beamforming vectors (e.g., at each iteration of Algorithms 1-3), we can directly extract the phase shift and force the amplitude to be one to get the desirable reflective beamforming vector with unit amplitudes. Based on this optimization principle together with alternating optimization for updating the coordinated transmit beamforming, we can obtain an efficient unit-amplitude reflection solution to the min-weighted-SINR maximization problem (P1). Fig. \ref{fig:antenna_number_unit_amplitude} shows the performance achieved by the three proposed algorithms with and without the unit-amplitude consideration. It is shown that each of the three proposed alternating optimization algorithms with the unit-amplitude reflection constraints achieves similar minimum SINR at users as the upper bound achieved by the counterpart without the unit-amplitude reflection constraints (in Algorithms 1-3), with a negligible performance loss. This shows that with slight modifications, our proposed algorithms are also applicable in the scenario when the reflection amplitude of the reflecting units is fixed. 

\section{Concluding Remarks}
In this paper, we investigated the IRS-aided multi-cell MISO system, with the objective of maximizing the minimum weighted SINR at all users by jointly optimizing the coordinated transmit beamforming vectors at the BSs and the reflective beamforming vector at the IRS, subject to the individual transmit power constraints at the BSs and the reflection constraints at the IRS. We proposed three different alternating-optimization-based approaches, namely exact-alternating-optimization (i.e., Algorithm 1), inexact-alternating-optimization (i.e., Algorithm 2), and low-complexity inexact-alternating-optimization (i.e., Algorithm 3), respectively, to solve the min-weighted-SINR maximization problem, by  balancing between the performance and complexity. Numerical results demonstrated that the dedicatedly deployed IRS considerably improves the SINR performance of the multi-cell MISO system by not only enhancing the received signal strength but also suppressing the inter-cell interference, especially for cell-edge users. It was also shown that the inexact-alternating-optimization approach is an efficient technique for jointly optimizing the transmit and reflective beamforming vectors with reduced complexity and guaranteed convergence, which outperforms the conventionally adopted exact-alternating-optimization approach. Due to the space limitation, there are some important issues unaddressed in this paper, which are briefly discussed in the following and will be our future work.

{\it Channel Estimation and Signaling Overhead Issue}: In this paper, we assumed perfect CSI available globally and ignored the CSI signaling overhead among BSs/IRS for simplifying the analysis and gaining the essential design insights. However, due to the passive nature of the IRS, the CSI between the IRS and BSs and/or between the IRS and users is difficult to obtain, and may yield a huge channel training overhead and require a sufficiently long channel estimation time, especially when the number of reflecting units at the IRS and the numbers of cells and users are large. Furthermore, due to the latency caused in the backhaul/fronthaul links, the delay of the CSI sharing among BSs/IRS may also introduce certain CSI imperfection. As a result, it is important to design efficient channel estimation and signaling 
methods to balance the trade-off between the CSI accuracy and training/signaling overhead for maximizing the performance gain for the IRS-aided multi-cell transmission.

{\it Hardware Impairments}:
This paper assumed continuous amplitudes and phase shifts for the IRS, which can be independently controlled to provide the maximum design flexibility. This, however, may increase the fabrication cost and the implementation complexity. To overcome this issue, alternative hardware designs with low-complexity amplitude and phase control (e.g., with limited resolutions) need to be considered \cite{IRS_wu}.  Furthermore, the amplitudes and phases between neighboring IRS units may be correlated in practice, instead of completely independent \cite{amp_phase_relat}. By taking into account these hardware impairment issues, how to design the IRS-aided multi-cell systems is an interesting problem for further investigation.

\end{document}